\documentclass[iicol, pdflatex, sn-basic]{sn-jnl}

\usepackage{graphicx}%
\usepackage{subcaption}
\usepackage{multirow}%
\usepackage{amsmath,amssymb,amsfonts}%
\usepackage{amsthm}%
\usepackage{mathrsfs}%
\usepackage[title]{appendix}%
\usepackage[dvipsnames]{xcolor}%
\usepackage{textcomp}%
\usepackage{manyfoot}%
\usepackage{booktabs}%
\usepackage{algorithm}%
\usepackage{algorithmicx}%
\usepackage{algpseudocode}%
\usepackage{listings}%
\usepackage{siunitx}%
\usepackage{lmodern}

\DeclareSIUnit\erg{erg}

\theoremstyle{thmstyleone}%
\theoremstyle{thmstyletwo}%

\theoremstyle{thmstylethree}%

\newcommand{\gc}{$\gamma$\,Cas}

\begin{document}

\title[Accreting companions of Be/\gc\ stars]{High energy emission powered by accreting companions of Be/\gc\ stars}

\author*[1]{\fnm{Rina G.}\sur{Rast}}\email{krast@uwo.ca}
\author[2]{\fnm{Ya\"el}\sur{Naz\'e}}
\author[3,7]{\fnm{Jonathan}\sur{Labadie-Bartz}}
\author[1]{\fnm{Carol E.}\sur{Jones}}
\author[4]{\fnm{Christiana}\sur{Erba}}
\author[5]{\fnm{Ken}\sur{Gayley}}
\author[6]{\fnm{Asif}\sur{ud-Doula}}
\author[3]{\fnm{Coralie}\sur{Neiner}}
\author[8]{\fnm{Jeremy J.}\sur{Drake}}

\affil*[1]{
    \orgdiv{Department of Physics and Astronomy}, 
    \orgname{The University of Western Ontario},     \orgaddress{\city{London}, \postcode{N6A 3K7}, \state{ON}, \country{Canada}}}

\affil[2]{
    \orgdiv{STAR Institute}, 
    \orgname{Universit\'e de Li\`ege},
    \orgaddress{\street{All\'ee du 6 Ao\^ut, 19c, B\^at B5C}, \city{Li\`ege}, \postcode{4000}, \country{Belgium}}}
    
\affil[3]{
    \orgdiv{LIRA, Paris Observatory}, 
    \orgname{PSL University, CNRS, Sorbonne University, Universit\'e Paris Cit\'e, CY Cergy University}, 
    \orgaddress{\street{5 place Jules Janssen}, \postcode{92195}  \city{Meudon}, \country{France}}}

\affil[4]{
    \orgname{Space Telescope Science Institute}, \orgaddress{\street{3700 San Martin Drive}, \city{Baltimore}, \postcode{21218}, \state{MD}, \country{USA}}}

\affil[5]{
    \orgdiv{Department of Physics \& Astronomy}, 
    \orgname{University of Iowa}, 
    \orgaddress{\street{203 Van Allen Hall}, \city{Iowa City}, \postcode{52242}, \state{IA}, \country{USA}}}

\affil[6]{
    \orgname{Penn State Scranton}, 
    \orgaddress{\street{120 Ridge View Drive}, \city{Dunmore}, \postcode{18512}, \state{PA}, \country{USA}}}   

\affil[7]{
    \orgdiv{DTU Space}, 
    \orgname{Technical University of Denmark}, 
    \orgaddress{\street{Elektrovej 327}, \city{Kgs., Lyngby}, \postcode{2800}, \country{Denmark}}} 

\affil[8]{\orgdiv{Advanced Technology Center}, \orgname{Lockheed Martin}, \orgaddress{\street{3251 Hanover St}, \city{Palo Alto}, \postcode{94304}, \state{CA}, \country{USA}}}  

\onecolumn
\abstract{The origin of the hard, bright X-ray emission that defines the \gc\ analog class of Be stars remains an outstanding question in Be star literature. This work explores the possibility that the X-ray flux is produced by accretion onto a white dwarf companion. We use three-dimensional smoothed particle hydrodynamics simulations to model the prototype \gc\ system assuming a white dwarf companion and investigate the accretion of the circumstellar material by the secondary star. We contrast these results to a model for 59\,Cyg, a  non-\gc\ Be star system with a stripped companion. We find that the secondary stars in both systems form disk-like accretion structures with Keplerian characteristics, similar to those seen in the Be decretion disks. We also find that white dwarf accretion can produce X-ray fluxes that are consistent with the observed values for \gc, while the predicted X-ray luminosities are significantly lower for the non-degenerate companion in 59\,Cyg. In addition, using the three-dimensional radiative transfer code, \textsc{hdust}, we find that these models produce H$\alpha$ emission consistent with the observations for both \gc\ and 59\,Cyg, and that the predicted polarization degrees across optical and UV wavelengths are at detectable levels. Finally, we discuss the impact that future UV spectropolarimetry missions could have on our understanding of these systems.}


\maketitle
\twocolumn
 
\section{Introduction}

More than a century ago, the discovery of emission lines in stellar spectra led to the definition of the Be spectral classification. Today, classical Be stars are understood to be massive, rapidly-rotating stars surrounded by decretion disks that are born through successive, transient ejections from the stellar surface at the equator \citep[][and references therein]{riv13}. While a complete description of the disk-building process has yet to be offered, recent work suggests that mechanisms such as nonradial pulsations and rapid rotation may play a role \citep{baa20, nei20, lab22}, and the mass ejection itself may be restricted to localized regions of the stellar surface \citep{lab25}. Be stars are dynamic and their observed spectral lines, photometric brightness, and polarization levels vary over timescales of hours to decades, making them appealing testbeds for stellar and disk physics.

Be stars are frequently found in binary systems, although the Be binary fraction is not well constrained. Their companions are usually in advanced evolutionary stages and can be detected by their UV emission, as is the case with stripped helium stars \citep{wang2021}, or by their bright X-ray emission, as is the case with neutron stars and white dwarfs (WDs)  \citep{reig2011, gau24}. The predominance of evolved companions has led to the idea that many Be stars may be the products of interacting binaries, where the initially lower-mass star in the binary is ``spun up" as it accretes matter and angular momentum from its evolving companion.

While Be stars are typically X-ray bright (see Sect.~\ref{subsec:Xray_summary}), those that emit atypically hard X-ray emission have been classified as ``\gc\ analogs'' after the archetype in which such features were first observed (for an observational history of \gc, see \citealt{har02}). The X-ray emission from these objects is associated with very hot plasma ($kT>5$\,keV) \citep{naz20}. The origin of this peculiarity is currently debated and may hold clues regarding the role of binary interaction in the Be phenomenon. Some have suggested that the X-rays could arise from small-scale star-disk magnetic interactions \citep{smi16} or from accretion onto a WD \citep{mur86, ham16, tsu18}. Investigations of these stars have focused on the optical and X-ray regimes, yet there are many outstanding questions.

The main goal of this work is to explore how UV observations of \gc\ objects may provide information about the origin of their X-ray emission and the nature of their companion stars. We begin with a review of previous X-ray (Sect.~\ref{subsec:Xray_summary}) and UV (Sect.~\ref{subsec:UV_summary}) studies of Be stars, and of studies of WD accretion (Sect.~\ref{subsec:binary_summary}) that have been linked to normal Be stars. We then present models of Be binary systems along with predicted observables to explore the origin of the X-ray and UV emissions in \gc\ analogs from a computational perspective, and compare these to past observations. We use smoothed particle hydrodynamics (\textsc{sph}) simulations to compare star-disk interactions in \gc\ to other Be binaries, adopting mass accretion rates from these \textsc{sph} simulations to predict the X-ray fluxes that could be produced by the systems. Then, we employ the radiative transfer code \textsc{hdust} \citep{car06, car08} to predict the H$\alpha$ emission lines and polarization across the V- and UV- bands that could be produced by the disks. The methods used in our simulations are presented in Sect.~\ref{sec:simulation_methods}. The disk structures produced by the \textsc{sph} simulations are detailed in Sect.~\ref{sec:sph_results}, while the predicted observables are explored in Sect.~\ref{sec:hdust_results}. Finally, in Sect.~\ref{sec:discussion_conclusions}, we summarize and discuss our results within the context of the new UV missions and instruments that have been proposed for future development, such as the small UV explorer mission concept {\em Polstar} \citep{paulNew} and the proposed UV spectropolarimeter {\em Pollux} \citep{Pollux2024} for the Habitable Worlds Observatory (HWO, \citealt{nas21}).

\subsection{An X-ray perspective of Be stars}
\label{subsec:Xray_summary}

In order to contextualize the predicted X-ray fluxes that follow, some background information on past X-ray studies of Be stars is required and is presented here. Of all high-energy domains, X-rays represent the most-studied wavelength range for Be stars. Be binaries with neutron star companions are easily detected as X-ray binaries as a result of their bright X-ray emission, and these systems represent a large fraction of all high-mass X-ray binaries \citep{reig2011, for23}. Since classical Be stars were not expected to be any brighter than normal B type stars at high-energy wavelengths, the discovery of moderate X-rays in \gc\ was noteworthy \citep{jer76, mas76}. This detection was confirmed and refined over the subsequent decades by several X-ray facilities (for a review, see \citealt{smi16}). The X-ray emission is thermal \citep{shr15} and moderately bright ($L_{\rm{X}}\sim 10^{-6}L_{\rm{BOL}}$), with a very high temperature of $kT\sim 12-14$ keV, and is highly variable on short timescales (seconds to hours) \citep{smi04, lop10, smi16}. The spectrum shows weak forbidden lines in He-like triplets, indicating the hot plasma has a high density or may be close to a UV source \citep{smi16}. Fluorescence lines are present, most notably in Fe at 6.4 keV \citep{lop10}, which suggests the presence of cool material near the hot plasma. Such properties are clearly at odds with X-ray characteristics of ``normal" OB stars, which display fainter ($L_{\rm{X}}\sim 10^{-7}L_{\rm{BOL}}$), constant, and soft ($kT\sim 0.6$keV) X-ray emission without any fluorescence component (for a review, see \citealt{rau22a}).

It is natural to consider whether \gc\ is an exceptional star or a member of a group of stars with these characteristics (see e.g. \citealt{pet82b}). Over the years, several other Be stars were indeed found to share similar X-ray properties, leading to the definition of the \gc\ phenomenon. About two dozen \gc\ analogs are now known, with an incidence rate of about 10\% in the whole Be population; the other, non-\gc\ Be stars display only soft and faint X-rays \citep{naz18, naz23}. 

The variability of \gc\ analogs is not limited to short timescales, as changes over months to years have also been observed \citep{smi16}. In this context, the prototype, \gc, is of course the most studied case. Several changes were found to be directly related to simultaneous variations in optical and UV (see Sect.~\ref{subsec:UV_summary}). In addition, spectral hardening events known as soft ``dips," caused by increased absorption have also been reported \citep{ham16, smi19b, rau22b}. The impacts of disk evolution on X-ray observations were also investigated, with very different results. In fact, \gc\ characteristics may be seen even if the disk emission is faint \citep{naz22b}. In addition, orbital phase or H$\alpha$ properties do not seem to play a role in the X-ray emission properties (e.g. \gc, $\pi$\,Aqr, $\zeta$\,Tau; see \citealt{naz25} and references therein). $\pi$\,Aqr did display a decrease of hardness and brightness the last time its disk dissipated, but at no time did it lose its \gc\ characteristics \citep{naz22b}. In fact, while no Be star has yet been seen to ``become" a \gc\ star, only one \gc\ star lost its defining characteristics: HD~45314. The hottest star amongst \gc\ analogs has made this transition as its disk  dissipated \citep{rau18, naz23}. Finally, most non-\gc\ Be stars have also been reported to vary (e.g. $\lambda$\,Eri, \citealt{smi97}) although the characteristics differ from \gc\ variability and the origin of the changes remains unknown, with possible explanations being contamination by a nearby source, or true variation of the Be star.

As mentioned previously, the origin of the peculiar X-rays from \gc\ analogs remains to be shown. Scenarios involving neutron star accretion during a propeller phase or disk-wind collisions could be excluded (\citealt{rau24, naz22a} and references therein). Two main scenarios persist, one involving accretion onto a WD \citep{mur86} and one requiring star-disk interactions (\citealt{smi16} and references therein). Up to now, they could not be distinguished in the X-ray range. 

Recently, several bright and very soft X-ray sources were found to be associated with Be counterparts in the optical range \citep{ken21, gau24, mar25}. Since the X-ray properties are characteristic of nova-like behaviour, it is possible that the companion stars in these systems are WDs. Unfortunately, no Be star has been seen to transition to such a state in the X-ray range, nor have nova-like cases been observed outside of their bright events. A direct link between such systems and \gc\ analogs (or other Be stars in general), therefore, still needs to be established or ruled out.

\subsection{Previous UV observations on Be and \gc\ analogs} \label{subsec:UV_summary}

To place our predicted observations in context, we provide a summary of past UV observations of Be stars. Since UV spectroscopy is a key diagnostic for the winds of hot stars, previous UV studies of Be stars in general (and \gc\ analogs in particular) have focused on wind lines (\citealt{gra87,pri89,sle94} and references therein). While the focus of this work is on Be star disks rather than their winds, we include brief background on wind-related studies for completeness. Historical studies revealed that wind indicators, including the spectral lines Si\,{\sc iv} and C\,{\sc iv}, are more prominent in early-type Be stars than in non-emission stars of the same spectral types \citep{sno81, gra87}. They also suggested a latitude dependence on the strength and variability of the winds \citep{gra87}. Recent simulations have indicated that such winds could ablate the disk of an early-type Be star \citep{kee16,kee18a,kee18b}.

Previous studies have also used UV observations to probe the structure of circumstellar environments near the stellar surface. \citet{gra87} and \citet{pri89} found that more than half the Be stars in their samples exhibited narrow, blue-shifted  absorption in N\,{\sc v}, Si\,{\sc iv} and C\,{\sc iv} which had stable velocities (see also \citealt{hen83,hen86}). These absorption features are variable, with reported lifetimes between one week and one month \citep{doa87}. \citet{cra00} noted similar blueshifted absorption features for \gc\ and evaluated the density of the associated material as intermediate between that of
the disk and that of the polar wind. These narrow absorptions seem to correlate with near-equator viewing of the Be stars \citep{gra87} and they appear much more often (and are stronger) when the violet peak of the double-peaked disk emission dominates (i.e. $V/R>1$, \citealt{doa85,doa87,doa89,tel94} and references therein). This suggests a link between the disk and other circumstellar material.

Additional features, with evolving radial velocities, have been interpreted as the usual corotating structures in front of the star through which the wind moves \citep{cra00}. Adding to the complexity of the UV observations, \citet{smi99} reported additional  narrow and faint absorption shifting from blue to red (``migrating subfeatures"), broader but stationary absorption varying over time, and sharp and nearly stationary absorption also appearing/disappearing over time in \gc. The UV spectra of Be stars, then, seem to present a large range of features. It remains to be examined how specific these features are to Be stars in general, or \gc\ objects in particular. 

Detailed surveys of Be stars combining UV wavelengths with optical or X-ray have tracked disk evolution through building phases and dissipation events \citep{doa85,gra87b}. \citet{daw84} monitored the Be binary and \gc\ analog $\zeta$\,Tau at UV and visible wavelengths, finding that the Si\,{\sc iii}, Si\,{\sc iv} and C\,{\sc iv} lines are formed in hotter, more ionized regions while lines such as N\,{\sc i} may be formed in the cooler regions of the disk. These studies highlight the utility of combining observations at different wavelengths to distinguish layers in the circumstellar regions of Be stars.

The more extensive efforts, however, focused on the bright star \gc. The first reports of behavior correlations at various wavelengths concerned specific events, whose rarity or abundance remains unknown to this day. A 64-hr Copernicus dataset with simultaneous H$\alpha$ monitoring revealed a short increase in H$\alpha$ correlated with small changes in some UV lines \citep{sle78} and an increase in X-ray emission \citep{pet82a}. In addition, there were three long UV observing campaigns involving \gc: 44 hr in January 1982 by the International Ultraviolet Explorer (IUE) at an orbital phase close to conjunction with the Be in front, 33 hr in January 1996 by IUE at a similar phase, and 21 hr in March 1996 by Hubble Space Telescope (HST) and Goddard High Resolution Spectrograph (GHRS) simultaneously with Rossi X-ray Timing Explorer (RXTE) at a quadrature phase. The GHRS data revealed ``dips" with amplitudes of $\sim$1\% in the UV continuum emission and timescales of hours \citep{smi98a}. Since such timescales are too short to be attributed to starspots, \citet{smi98b} suggested they may be evidence for clouds attached to the stellar surface and extending to several tenths of a stellar radius, but this suggestion has not been confirmed.

In addition, X-ray fluxes and UV continuum fluxes were found to anti-correlate \citep{smi98a}, and Si absorption weakened while Fe\,{\sc v} absorption strengthened when X-ray flux increased, which could all suggest a change in ionization due to X-rays \citep{smi98b,smi03}. By contrast, the cycles of about 70 days, seen at both optical and X-ray wavelengths, are not seen in the UV \citep{smi03}, perhaps because they are related to the disk rather than the Be star.

Long-term monitoring of a large sample of Be stars that includes \gc\ and non-\gc\ objects is lacking, so potential differences remain largely unexplored. A new UV spectropolarimetric mission, such as {\em Polstar} \citep{paulNew}, {\em Pollux} on the HWO \citep{Pollux2024}, or {\em Arago} \citep{mus23}, can fill this gap, boosting our understanding of disk phenomena and their relation to \gc\ analogs. 

\subsection{Previous UV observations on accreting WDs, and potential links to Be+WD binaries} \label{subsec:binary_summary}

Since this work is focused on the possibility that WD accretion could be powering the X-ray luminosity of \gc\ analogs, it is useful to draw comparisons to other X-ray bright systems with WD components. Here, we provide a brief summary from the literature describing cataclysmic variables (CVs) and related systems, which host accreting WDs in very close orbits around low-mass, near main sequence (MS) stars, to elucidate similarities and differences from the systems studied here. 

CVs have been observed to go through so-called high, low, and intermediate states. During the high state, the accretion disk greatly outshines the WD photosphere. At other times, the WD photosphere dominates the UV flux with only a small contribution from an accretion disk, resulting in a low state. Stars observed between these two extremes are said to be in the intermediate state. The physical size of the accretion disks formed around the WD are quite small, $\sim$0.3 R$_{\odot}$ or less. For the well-studied nova-like system MV Lyrae (usually in a high state, but occasionally dropping to a low state; \citealt{god2017}), the pure WD atmosphere typically exhibits a UV flux level at 1500 \AA \, of about 2$\times10^{-14}$ erg/s/cm$^{2}$/\AA. In its intermediate state, the flux level rises by about a factor of 5, and the accretion disk can be described by a model with a mass accretion rate of 2.4$\times10^{-9}$ M$_{\odot}$ yr$^{-1}$.

The cool MS star in CV systems, for which typical effective temperatures ($T_{\rm{eff}}$) are smaller than 4000 K \citep{kin89}, contributes negligibly in the UV. This is not the case for Be star binaries, where the hot ($T_{\rm{eff}}$ $\gtrsim$ 10,000 K as per \citealt{cox00}) and relatively large (radii exceeding 2.7 R$_{\odot}$, \citealt{cox00}) B-type primary is a strong source of UV flux. For example, a typical B1V star has $T_{\rm{eff}} = 26200$ K and $R = 7$ R$ _{\odot}$. Comparing the luminosity of the CV disk in the intermediate state for MV Lyrae with a typical B1V star with the Stefan-Boltzmann relation,
\begin{equation}
\frac{L_{\rm CV disk}}{L_{\rm B1V}} = \left( \frac{R_{\rm CV disk}}{R_{\rm B1V}} \right)^2 \times \left( \frac{T_{\rm CV disk}}{T_{\rm B1V}}\right)^4 \,,
\end{equation}
puts the B1V star as being about 25000 times more luminous. Thus, a typical CV disk would be all but invisible in a scenario with an early-type Be primary and a WD companion accreting material from the large Be star disk. 

However, there is no reason to suspect that a putative WD companion orbiting a Be star would possess an accretion disk that is qualitatively similar to CV systems. The size of the WD accretion disks in CV systems is set by the orbital separation ($a$), where $R_{\rm disk, outer} = a/3$. For MY Lyr, $a$ is approximately 1.2 R$_{\odot}$ and the orbital period is only 3.2 hours.  By contrast, Be binaries have typical orbital periods of between about 30 and 300 days. For example, the binary separation in 59\,Cyg (P$_{\rm orb}$ = 28.2 d) is about 75 R$_{\odot}$ \citep{pet13}, and in $\gamma$ Cas (P$_{\rm orb}$ = 206.3 d) is about 350 R$_{\odot}$ \citep[e.g.][]{baa23}. This allows for the possibility of significantly larger WD accretion disks in Be binaries. This effect is seen in our simulations detailed below, where the disk around the secondary in \gc\ extends past $\sim$25 R$_\odot$ while the circumbinary structure in 59\,Cyg is no larger than $\sim$10 R$_\odot$.

The temperature of a presumed WD accretion disk in a Be binary also has the potential to be higher than in CV systems. The temperature structure of Be disks is set primarily by the density profile within the disk and the incident flux from the Be star. On average, the Be disk temperature is often described to be about 60\% of the $T_{\rm eff}$ of the Be star (see Sect.~\ref{sec:simulation_methods}). If we continue using the example of a B1V primary, this would correspond to about 16,000 K. Presumably, material at this temperature would only get hotter in the process of accreting onto a WD. We note that if the secondary is not coplanar with the Be star disk, the disk may tilt away from the Be star's equator and its average temperature may be larger than in coplanar disks, as noted by \citet{suf23} when studying tilted Be star disks.

In summary, accretion disks in a Be+WD binary have the potential to be both significantly larger and hotter than in CV systems. However, WD accretion disks have never been reported in Be star systems, although there are a few reports of accretion disks around hot subdwarf companions to Be stars \citep[e.g.][]{cho2018}. 

It is useful to also consider the time domain. In typical Be binaries, the Be star is about ten times the mass of its companion. Thus, the RV semi-amplitude of the low-mass companions is high compared to the RV motion of the Be star. Any emission features arising from a WD accretion disk will move in anti-phase, and with a considerably higher velocity range, compared to the RV motion of the Be primary. This provides a natural opportunity to disentangle signals from the Be decretion disk and the WD accretion disk.

Symbiotic systems, where the donor star is a red giant and the accretor is a compact object, may also show hard X-rays \citep{cha05, ken09, muk16}, but with longer orbital periods than CVs, possibly making them better proxies to \gc\ systems \citep{naz24}. The accretion may be wind-driven or through wind Roche lobe overflow \citep{pod07}. Accretion disks have been reported or proposed in some of these objects \citep{lop18, lun18, kum21}. When such an accretion disk is formed, it is larger than in CVs, as could be expected from the larger orbit \citep{dus86}. Some of these systems are powered purely by the accretion process, with accretion rates of the order of \SI{e-9} to \SI{e-11} M$_{\odot}$ yr$^{-1}$ \citep{sio19,kum21, lim24}, while others are shell-burning and produce emission through nuclear processes in addition to accretion \citep{lun13, muk16}.

Observations of symbiotic systems in the UV have enabled the direct detection of emission from the WD (e.g. \citealt{san10}) but International Ultraviolet Explorer data also revealed information about the geometry of outflow from the jets of accreting WDs \citep{tom88}. In addition to providing details about the nature of both binary components, UV spectra have suggested that such jets can be produced by radiative shock from a precessing accretion disk, as in the case of R Aqr \citep{mei95}. Systematic redshifts of the He\,\textsc{ii} 1640 \AA\ line have also been observed in CI Cyg, and interpreted as evidence for an asymmetric wind interaction shell, or an accretion disk wind \citep{mik06}. Based on emission line shifts in several symbiotic systems, \citet{fre10} similarly suggested the presence of an expanding P Cygni profile from the wind of the cooler star. \citet{mun89} proposed that the majority of the UV emission lines observed in symbiotic systems are produced by the atmosphere of the cooler star on the side being heated by the compact object, which complicates comparison with \gc\ analogs.

If a given Be star has a WD companion, is there any hope of detecting a WD accretion disk via characteristic UV emission lines? In the sections below, we begin to address this question by performing \textsc{sph} simulations of circumstellar material in Be binary systems in order to determine the mass flux into the accretion disk and its approximate physical properties.

\section{Methods} \label{sec:simulation_methods}

Our \textsc{sph} simulations implemented the 3D \textsc{sph} code developed by \citet{ben90a} and \citet{ben90b} and refined by \citet{bat95}. This tool was edited for the purpose of studying Be star systems by \citet{oka02}. It simulates the two components of a binary system as point mass sink particles, initially without a disk present in the system. In the first time-step, a shell of gas particles is injected at a radius of 1.04 stellar equatorial radii ($R_{\star}$) around the primary sink particle, which represents the Be star. We continue to inject 40,000 particles every $1/(2\pi)$ orbital periods for the entire simulation. The particles are launched with a Keplerian velocity and with enough angular momentum to orbit at the injection radius. However, the majority of these particles lose their angular momentum through interactions with each other and are accreted by the primary star. Viscous torques allow the surviving particles to move outward to form a disk. The shear viscosity, $\nu$, of the disk is given by, 
\begin{equation}
    \nu= \frac{1}{10}\alpha_{\rm{SPH}} c_s h \,,
	\label{eq:ss_viscosity}
\end{equation}
where $h$ is the smoothing length and  $\alpha_{\rm{SPH}}$ is the linear artificial \textsc{sph} viscosity scaling parameter \citep{mon83, oka02}. Here, $c_s = \left( kT/\mu m_{\rm{H}} \right)^{1/2}$ represents the disk's isothermal sound speed where $k$ is the Boltzmann constant, $T$ is the disk temperature, $\mu$ is the mean molecular weight of the gas, and $m_{\rm{H}}$ is the hydrogen mass. We set $\alpha_{\rm{SPH}}$ to 1, which is roughly equivalent to the value for the Shakura–Sunyaev viscosity parameter $\alpha_{\rm{SS}}$ = 0.1 used by e.g. \citet{suf22}, \citet{suf25}, \citet{ras25}, and which is consistent with observation-based estimates for Be star disk viscosities \citep{rim18, gho21, mar21}. Setting $\alpha_{\rm{SPH}}$ to a constant value allows us to maintain a fixed viscosity throughout the disk, without the scale height dependence introduced by mimicking the Shakura–Sunyaev parameter as seen in other works. The scale height $H$ of a thin, isothermal Be star disk in vertical hydrostatic equilibrium can be defined as 
\begin{equation}
    H = \frac{c_s R_{\star}}{v_{\rm{orb}}} \left( \frac{r}{R_{\star}} \right)^{3/2} \,,
    \label{eq:scale_height}
\end{equation}
where $c_s$ is the sound speed in the disk as defined above, $v_{\rm{orb}}=\left( GM_\star / R_{\star} \right)^{1/2}$ is the circular Keplerian orbital velocity at the stellar equator and $r$ is the radial coordinate in the disk \citep{lig74, car08}. This expression holds while $H \ll r$ but is not applicable in the outer, flared regions of the disk. Since the scale height of the disk deviates from its theoretical proportionality $h \propto r^{3/2}$ near the secondary star, especially if an accretion disk is formed, we choose to define the viscosity independently of scale height. Therefore, keeping $\alpha_{\rm{SPH}}$ fixed allowed us to accurately represent the behavior of particles near the secondary star. 

The volume density distribution $\rho(r,z)$ of a Be star’s disk can be expressed in terms of the scale height $H$ as 
\begin{equation}
    \rho(r,z) = \rho_0 \left( \frac{R_{\star}}{r} \right)^{n} \exp \left( \frac{-z^2}{2H^2}\right) \,,
    \label{eq:vol_dens}
\end{equation}
where $r$ is the radial coordinate as defined above, $z$ is the vertical position in the disk, $n = 3.5$ for a steady-state, isothermal disk and $\rho_0$ is the density of the disk at $r=R_{\star}$ and $z=0$ \citep{bjo05}. As with the scale height, this expression assumes a steady-state disk in hydrostatic equilibrium. Similarly, the surface density is described using
\begin{equation}
    \Sigma(r) = \Sigma_0 \left( \frac{R_{\star}}{r} \right)^{m} \,,
    \label{eq:surf_dens}
\end{equation}
where $\Sigma_0$ is the surface density at $r=R_{\star}$ and $m$ takes a value of $n - 1.5$, or 2 for a steady-state disk \citep{bjo05}. 

Our simulations used particle splitting, a technique that increases the resolution in the outer regions of the disk where the density is lower and the particle count is limited. This method was implemented by \citet{rub25} and uses conditions consistent with the work of \citet{kit02}. It was used by \citet{ras25} in models of short-period Be/X-ray binaries. This technique achieves higher resolution by reducing the mass of individual particles and increasing the number of total particles, and its advantage lies in accomplishing this without over-resolving the dense inner disk. 

\begin{table}[ht]
    \caption{Parameters of the \textsc{sph} models.}
    \label{tab:sys_params}
    \centering
        \begin{tabular}{lcccr}
        \hline
        Parameter & \gc & 59\,Cyg\tnote{\it{d}} \\
        \hline
            Primary mass (M$_{\odot}$) & 13\tnote{\it{a}} & 7.89 \\
		Secondary mass (M$_{\odot}$) & 0.98\tnote{\it{a}} & 0.77 \\
		Primary radius, $R_{\star}$ (R$_{\odot}$) & 10\tnote{\it{b}} & 6.25 \\
		Secondary radius, $R_{2}$ (R$_{\odot}$) & 0.01\tnote{\it{c}} & 0.39 \\
            Primary $T_{\rm{eff}}$ (K) & 25000\tnote{\it{b}} & 21800 \\
            Orbital period (d) & 206.3\tnote{\it{a}} & 28.2 \\
		Eccentricity & 0\tnote{\it{a}} & 0.14 \\
        \hline
	$\alpha_{\rm{SPH}}$ & \multicolumn{2}{c}{1} \\
        Injection radius (R$_{\star}$) & \multicolumn{2}{c}{1.04} & \\
        Mass loss rate (M$_{\odot}$ yr$^{-1}$) & \multicolumn{2}{c}{ 1$\times$10$^{-8}$} \\
        \hline
        \end{tabular}
        \begin{tablenotes}
            \item[a] From \citet{nem12}.
            \item[b] From \citet{sig07}.
            \item[c] From \citet{par17}.
            \item[d] From \citet{pet13}.
        \end{tablenotes}
\end{table}

We chose to model real systems and selected two targets: \gc\ and 59\,Cyg. The former object is the prototype of the \gc\ category, so is a logical choice to represent this class of objects. We assumed that the companion in this system is a WD \citep{app02, gie23}. The latter object is a normal Be star paired with a hot, helium-burning subdwarf (sdO) companion \citep{pet13}. It provides a point of comparison to the accreting WD case. In addition, its orbit is slightly eccentric, allowing us to probe the effects of non-circular orbits in these systems. The parameters used in the \textsc{sph} simulations are shown in Table \ref{tab:sys_params}. The stellar masses and system parameters  for \gc\ were taken from \citet{sig07} and \citet{nem12}, with the radius of the secondary estimated from Fig. 9 of \citet{par17}. The parameters for 59\,Cyg were taken from \citet{pet13}. The disks were modeled as isothermal, with their respective temperatures set to 60\% of the Be star's $T_{\rm{eff}}$ \citep{mil99a, mil99b, car06}. The mass loss rate in both simulations was held constant at \SI{1e-8}{} M$_{\rm{\odot}}$ yr$^{-1}$, consistent with other recent works for ease of comparison \citep{cyr20, suf25, ras25, rub25}.

To provide a wider range of predicted observables from these systems, we used the output from our \textsc{sph} models to compute observables with the 3D nonlocal thermodynamic equilibrium (NLTE) Monte Carlo radiative transfer code, \textsc{hdust} \citep{car06, car08}. This process relied on an interface that converts the \textsc{sph} output at a given moment in the simulation to a grid of cells that contains information about the particle positions, velocities and densities. For recent examples of this interface coupled with \textsc{hdust}, see \citet{suf24} and \citet{suf25}. \textsc{hdust} produces observables using models for the stellar atmosphere and a dustless disk of hydrogen. The code simulates the paths of photons produced by the Be star as they travel through the disk to calculate the hydrogen ionization fractions and energy level populations, as well as the temperature structure of the disk \citep{car06}. This information is then used to compute observables such as the spectral energy distribution, H$\alpha$ profiles, and polarization levels, among others. For these systems, we focused on the predicted H$\alpha$ profiles as well as the polarization degree at visual and UV wavelengths.

We produced \textsc{hdust} simulations for 30 time-steps in the \textsc{sph} output for both \gc\ and 59\,Cyg, spanning five orbital periods. We used the same stellar parameters used in the \textsc{sph} simulations, shown in Table \ref{tab:sys_params}. We generated observables for inclinations, $i$, of 43$^\circ$ for \gc\ \citep{baa23} and 70$^\circ$ for 59\,Cyg \citep{mai05}. Additionally, we specified the rotation rate of each star as a fraction of critical, $W = v_{\rm{rot}}/v_{\rm{orb}}$, where $v_{\rm{rot}}$ is the stellar rotation rate and $v_{\rm{orb}}$ is the Keplerian circular orbital velocity at the equator, as defined above. The rotational values for both stars are uncertain \citep{baa23}. For \gc, we chose $W=0.9$ based on estimates for near-critical velocities suggested by \citet{cha01}, although it must be noted that it is difficult to reconcile high rotation rates with the inclination angle of 43$^{\circ}$. For 59\,Cyg, we used $v \sin{i}=379$ km s$^{-1}$ \citep{pet13} with an inclination of $i=70^{\circ}$ \citep{mai05}, yielding $W=0.73$. Simulations describing the effect of $W$ on the observables of Be stars and their disks are described in \citet{ras25b}.

\section{{\sc{sph}} models}
\label{sec:sph_results}

In this section, we present an analysis of the disks created in our simulations and compare their characteristics to the expected values for thin, Keplerian disks. This analysis forms a basis for an interpretation of the predicted observables that follow. 

We allowed the \textsc{sph} simulations to evolve with constant mass ejection while tracing the disk structure and measuring the accretion rates of the secondary stars. To estimate accretion rates, we counted the number of particles which entered a variable accretion radius centered on the secondary star. We set this radius to be 5\% of the Roche lobe, $r_{\rm{L}}$:
\begin{equation}
    r_{\rm{L}} = \frac{0.49q^{2/3}}{0.6q^{2/3} +\ln{(1+q}^{1/3})} D\,,
	\label{eq:roche_lobe}
\end{equation}
\noindent where $q$ is the ratio of the secondary stellar mass divided by the primary stellar mass and $D$ is the binary separation \citep{egg83}. We chose an accretion radius following \citet{rub25} for simulations based on the Be star, $\pi$\,Aqr, whose secondary star may be a WD \citep{mur86, heu24}. Due to the low sound speed in the disk, the Bondi radius defined by $r_{\rm{B}} = GM/c_{\rm{s}}^2$ \citep{fra02} was an unreasonably large value to use for this purpose. We experimented with reducing the size of the accretion radius to smaller than 5\% of the Roche lobe and found that it had no significant effect on the accretion rate in either of these systems. From Equation~\ref{eq:roche_lobe}, we can see that the Roche lobe and accretion radius are constant throughout an orbital period for circular orbits. For an eccentricity larger than zero, however, these parameters vary with the distance between the two stars. 

\begin{figure*}[!tbp]
  \begin{subfigure}[b]{\textwidth}
\includegraphics[width=\textwidth]{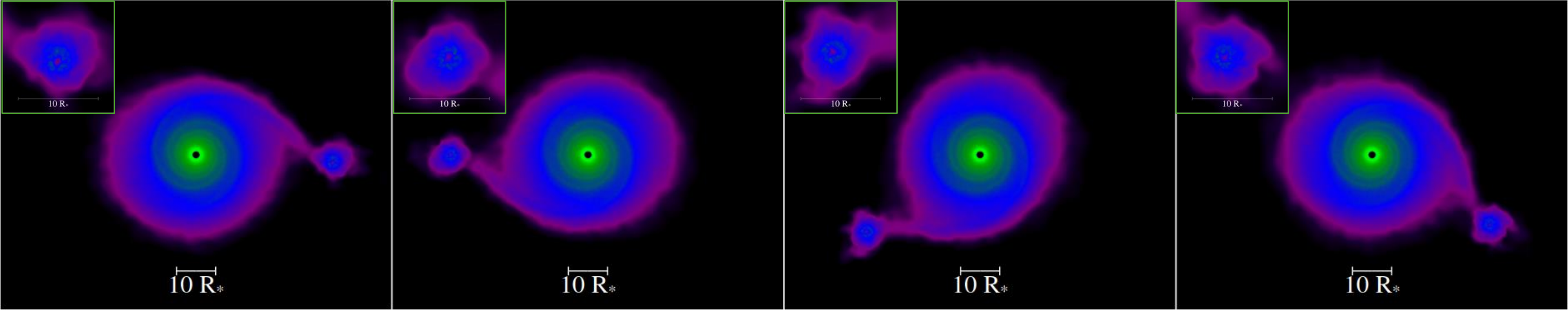}
  \end{subfigure}
  \hfill
  \begin{subfigure}[b]{\textwidth}
  \includegraphics[width=\textwidth]{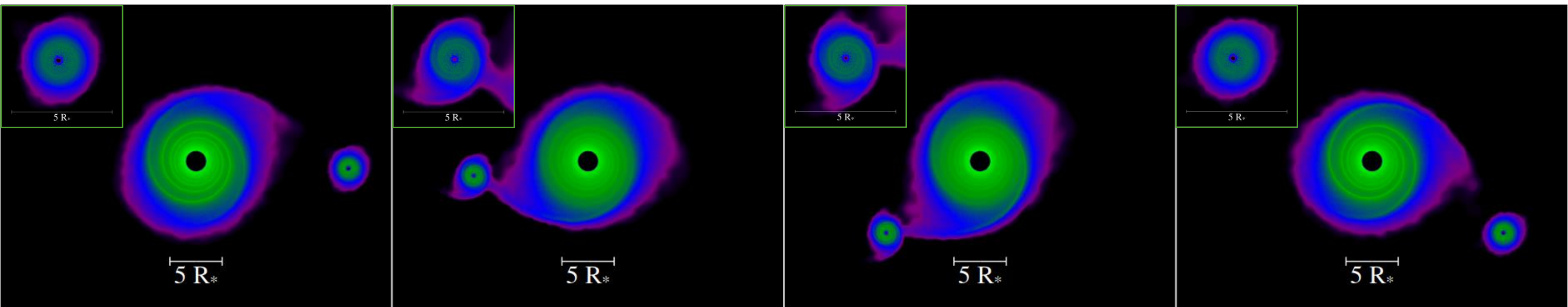}
  \end{subfigure}
  \caption{Top-down snapshots of the \textsc{sph} simulation for \gc\ (top) and 59\,Cyg (bottom) at 75.0 P$_{\rm{orb}}$, 75.5 P$_{\rm{orb}}$, 75.6 P$_{\rm{orb}}$ and 75.9 P$_{\rm{orb}}$ (from left to right). The insets at the top left of each image show an enlarged view of the circumsecondary region. As explained in the text, $R_{\star}$ represents the equatorial radius of the Be star in each system. Images rendered using \textsc{splash} \citep{pri07}.}
  \label{fig:splash}
\end{figure*}

Figure~\ref{fig:splash} shows snapshots of the simulations after 75 P$_{\rm{orb}}$ at different orbital phases, at which point the simulations had reached quasi steady-state behaviour. In addition to a decretion disk around the Be star, each of our models produces an accretion disk around the secondary star. Similar accretion disks were found in simulations by \citet{ras25} and \citet{rub25} using the same code and similar methods, as well as \citet{mar14} using the SPH code \textsc{phantom}. Thanks to our choice of small accretion radii, we are able to investigate regions of these accretion disks to very small distances from the secondary star, in addition to analyzing the primary star's more sizable disk. The use of particle splitting increased the resolution in the Roche lobe of the secondary star to thousands of particles for \gc\, and several tens of thousands in 59\,Cyg. This facilitates a comparison between the disks around the secondary in \gc\ and 59\,Cyg. Since the X-ray luminosities are powered by accretion from these disks, understanding the disk structure can provide information on how this accretion occurs. 

\begin{figure*}
    \centering
    \includegraphics[width=\linewidth]{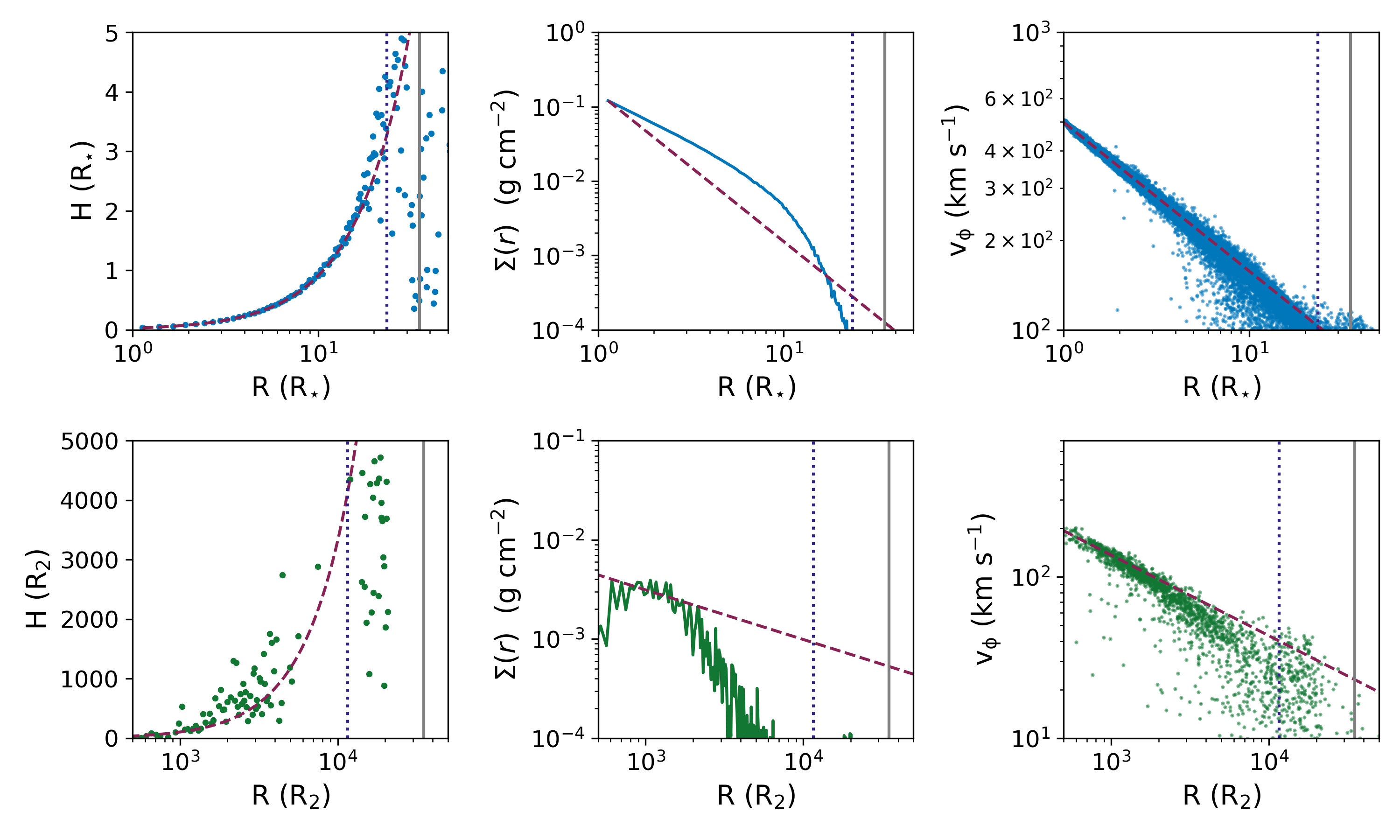}
    \caption{Disk scale heights (left), azimuthally averaged surface densities (center), and azimuthal velocities (right) for the \gc\ model. The top row shows these quantities for the system using a grid centered on the primary star, while the bottom row shows the values for a grid centered on the secondary star. The dashed line indicates the theoretical value for each quantity. The solid vertical line indicates the position of the secondary star (top row) or primary star (bottom row). The dotted line represents the L1 point. Only the particles with negative specific energies with respect to each star are shown.}
    \label{fig:gCas_scaleheights}
\end{figure*}

\begin{figure*}
    \centering
    \includegraphics[width=\linewidth]{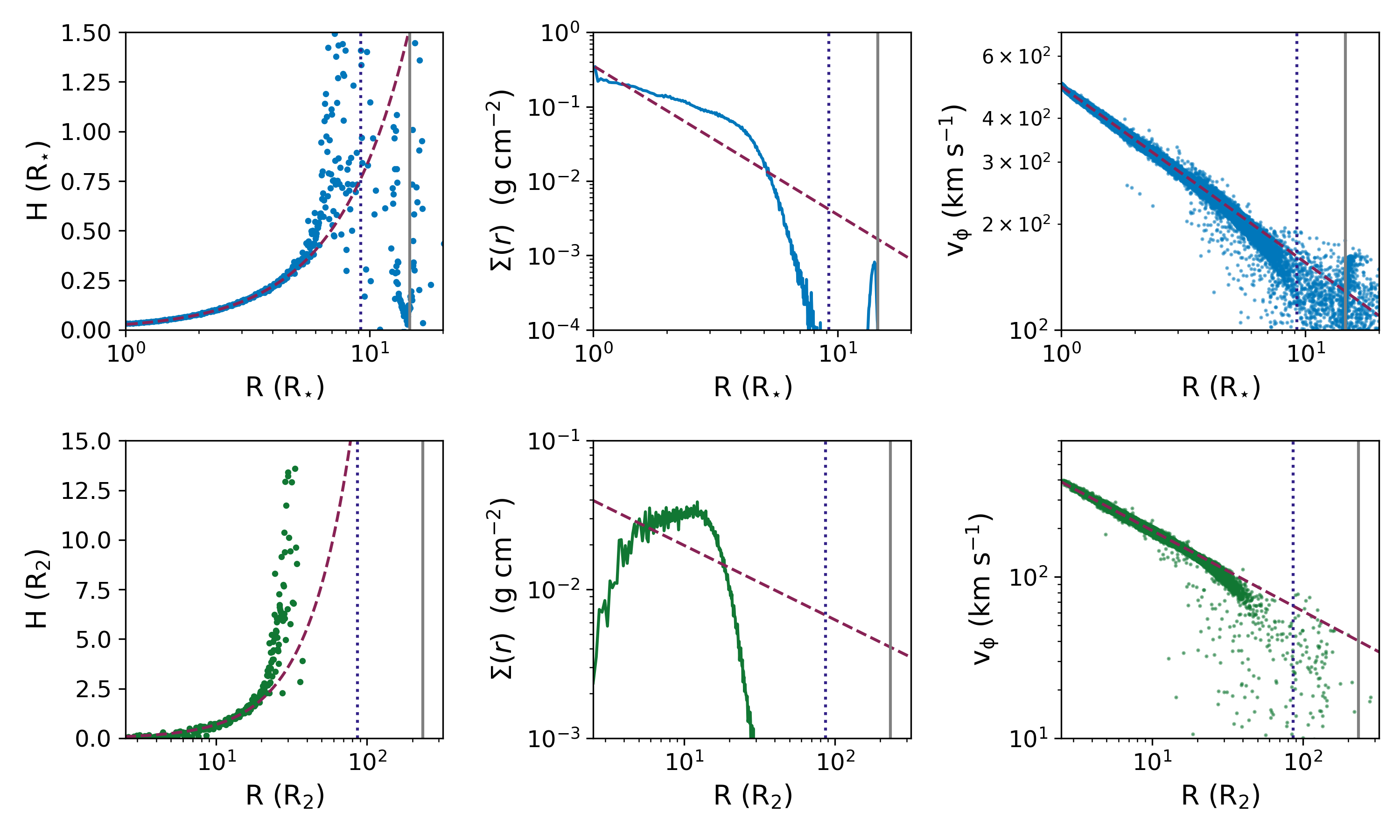}
    \caption{Same as Fig. \ref{fig:gCas_scaleheights}, but for 59\,Cyg.}
    \label{fig:fiftynine_cyg_scaleheights}
\end{figure*} 

The scale heights, azimuthally averaged surface densities, and azimuthal velocities, $v_{\phi}$, of the particles bound to the primary and secondary stars for \gc\ are shown in Fig.~\ref{fig:gCas_scaleheights}, while the same quantities are shown in Fig.~\ref{fig:fiftynine_cyg_scaleheights} for 59\,Cyg. We defined a particle as bound to a star if it had a negative specific energy with respect to that star. We note that by this definition, some particles at a given time step will be bound to the primary star at radial distances near the secondary; these particles are likely to either become bound to the secondary in a subsequent time step, leave the accretion disk entirely, or are located in diffuse regions of the disk not near the angular coordinate of the secondary. We calculated the scale heights by fitting a Gaussian to the volume density for slices of the disk along the radial direction, and compared them to the theoretical value predicted by Equation \ref{eq:scale_height}. The surface density was found by creating a grid centered on the star with the innermost edge at the stellar surface and integrating each grid cell vertically to find the azimuthally averaged density, similar to the method used by \citet{cyr17}. The theoretical values were found using Equation~\ref{eq:surf_dens}, using $m = 2$ for the decretion disk and $m = 0.5$ for the accretion disk. We also compare the measured azimuthal velocities to the theoretical Keplerian values given by $v_{\phi} = \sqrt{GM/r}$ where $M$ is the mass of the primary and secondary star for the decretion and accretion disk, respectively.

In the inner regions of the disks for both systems, Equation~\ref{eq:scale_height} provides an appropriate estimate of the actual scale heights in the decretion disk around the Be star, and a reasonable estimate for the accretion disk around the secondary. The measured scale heights follow the theoretical values as shown in the left panels of Fig.~\ref{fig:gCas_scaleheights} and Fig.~\ref{fig:fiftynine_cyg_scaleheights} in these regions. However, the scale heights in the outer regions deviate significantly, particularly in the model for 59\,Cyg. The trend departs entirely from the predicted values at distances larger than the first Lagrangian (L1) point. Since the decretion disk in \gc\ is much more radially extended than in 59\, Cyg, it flares significantly and we find much larger scale heights, as expected. 

The disk surface density, shown in the center columns of Fig. \ref{fig:gCas_scaleheights} and \ref{fig:fiftynine_cyg_scaleheights}, shows clear differences between primary and secondary stars. The rate of decline in the innermost regions of the decretion disks centered on the primary star is shallower than the predicted $n = 3.5$ for a steady-state disk. We can identify radial distances in both systems where the surface density of the decretion disk drops substantially. This represents the outer ``truncation radius," where the resonant torques prevent the outward flow of disk material \citep{oka02, pan16}. This radius has also been referred to as the ``transition radius" by \citet{suf22}, since a significant portion of the disk can extend beyond this point. More recently, high-resolution simulations by \citet{rub25} showed that the transition can span several R$_{\star}$ and varies azimuthally, following the Roche equipotentials. Within this transition radius, matter accumulates and leads to the increased surface density. The accretion disk centered on the secondary star in the \gc\ system roughly follows the theoretical drop-off prescribed by Equation~\ref{eq:surf_dens} with m = $0.5$ within the densest $\sim$2000 secondary radii ($R_2$), while the secondary’s disk in 59\,Cyg shows enhanced surface density with a slope shallower than $m = 0.5$ up to roughly 20 $R_2$, and drops off rapidly past this point.

\begin{figure}[!tbp]
  \begin{subfigure}[b]{0.5\textwidth}
    \includegraphics[width=\textwidth]{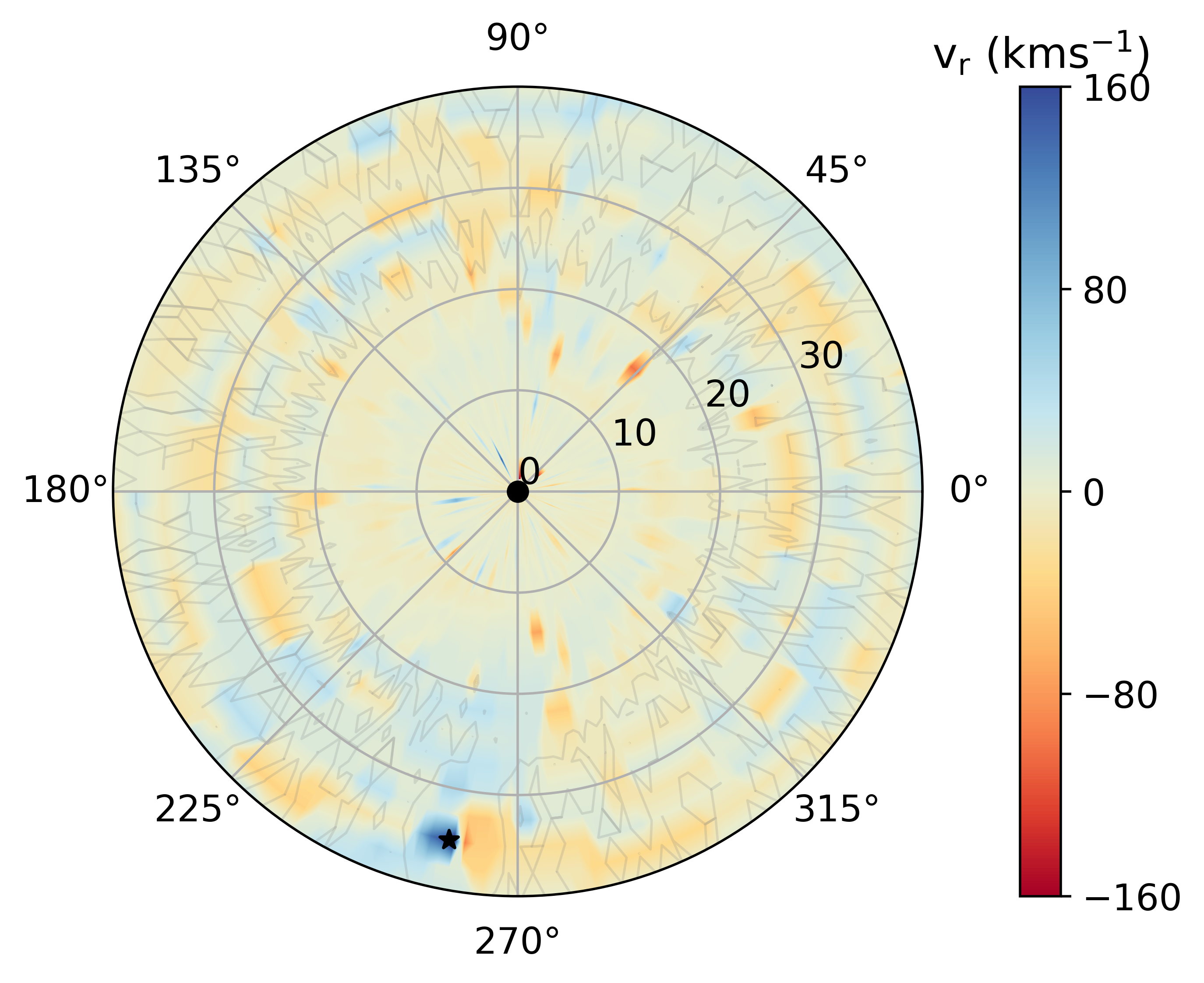}
  \end{subfigure}
  \hfill
  \begin{subfigure}[b]{0.5\textwidth}
  \includegraphics[width=\textwidth]{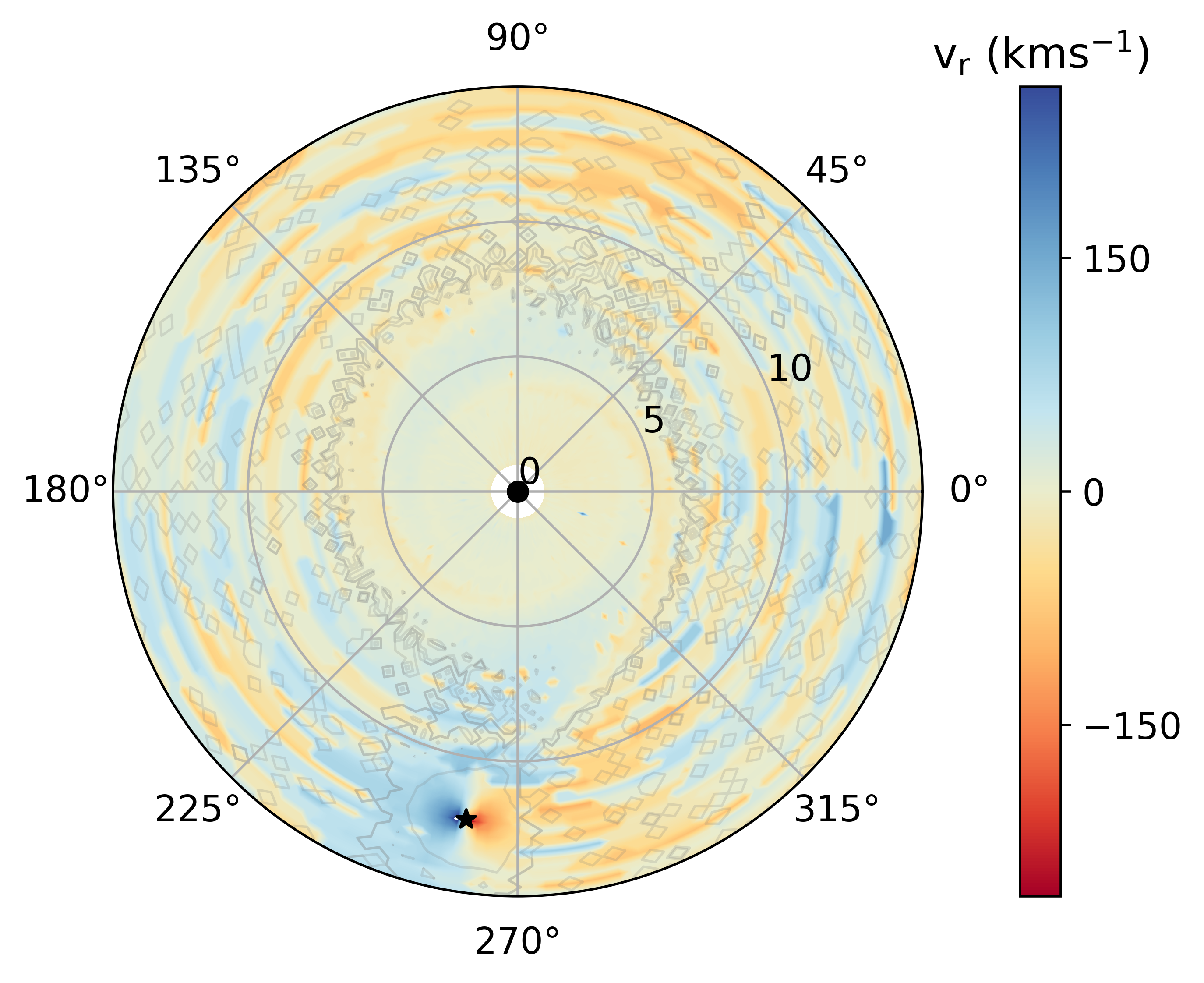}
  \end{subfigure}
  \caption{Radial velocity maps for \gc\ (top) and 59\,Cyg (bottom) models, measured with respect to the primary star. The dot at the center represents the primary star, while the secondary is represented by the black star and the contours represent surface density. The radial distances are in $R_{\star}$.}
  \label{fig:vrad_primary}
\end{figure}

\begin{figure}[!tbp]
  \begin{subfigure}[b]{0.5\textwidth}
    \includegraphics[width=\textwidth]{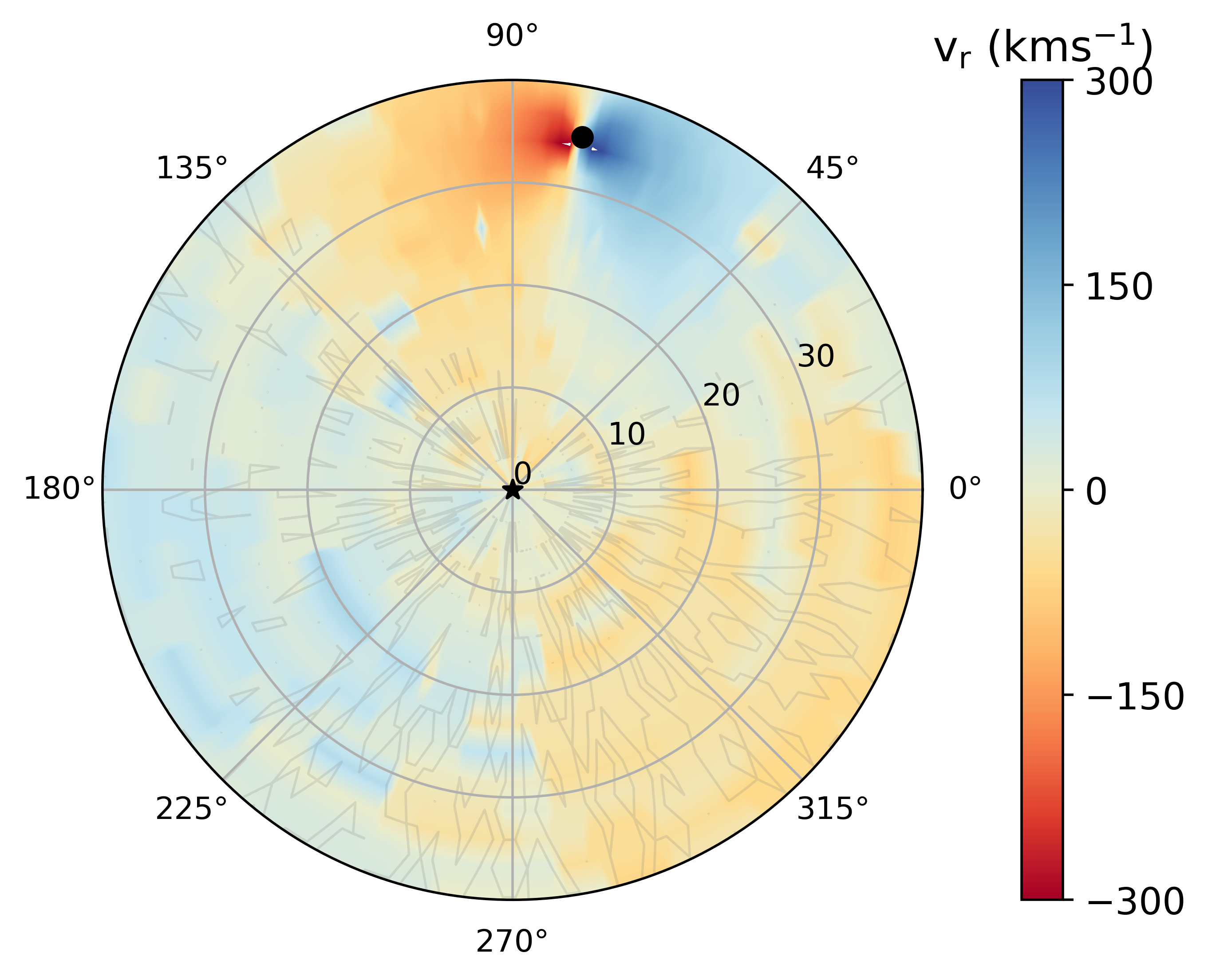}
  \end{subfigure}
  \hfill
  \begin{subfigure}[b]{0.5\textwidth}
  \includegraphics[width=\textwidth]{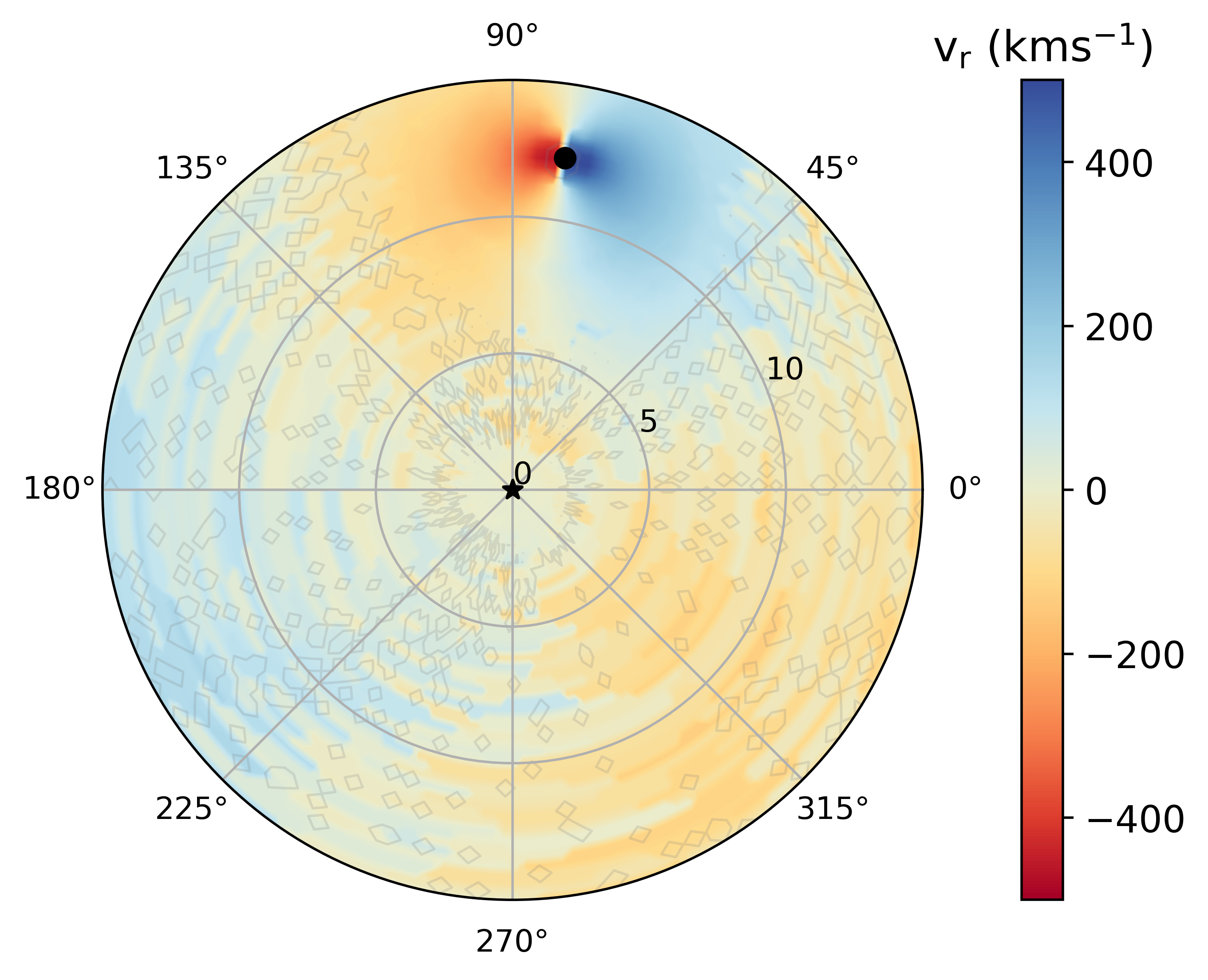}
  \end{subfigure}
  \caption{Same as in Figure~\ref{fig:vrad_primary}, but measured with respect to the secondary star, which is denoted as the black star at the center of each image.}
  \label{fig:vrad_secondary}
\end{figure}

The azimuthal velocities of the decretion and accretion disks in both systems, shown in the right panels of Figures~\ref{fig:gCas_scaleheights} and~\ref{fig:fiftynine_cyg_scaleheights}, closely follow Keplerian values. This is expected for a Be star disk, but is an important finding for the accretion disk around the secondary star. A radial velocity map measured with respect to the primary star is shown in the left and right panels of Fig.~\ref{fig:vrad_primary} for \gc\ and 59\,Cyg, respectively. The radial velocities for a quasi-Keplerian disk should be on the order of a few tens of km s$^{-1}$. We see that this is true in the inner disks for both primary stars, but the outer regions, especially around the secondary stars, show larger radial velocities that exceed the disk sound speed. Similarly, the radial velocities with respect to the secondary stars, shown in Fig.~\ref{fig:vrad_secondary}, are consistent with a Keplerian disk very close to the secondary and surpass the sound speed at large radii. 

It is important to note that unlike our \textsc{hdust} simulations, which find a self-consistent disk temperature, our \textsc{sph} simulations assume isothermality and presently do not have the capability to include radiative transfer. In addition, they do not include the effects of magnetism or stellar winds. Therefore, the structure of the disk surrounding the secondary star as described above is an exclusively hydrodynamical estimation. However, we can conclude that the secondary star in both systems builds a structure with characteristics consistent with an accretion disk. 

\section{Predicted observables} 
\label{sec:hdust_results}

\begin{figure}\centering
\subfloat{\includegraphics[width=\linewidth]{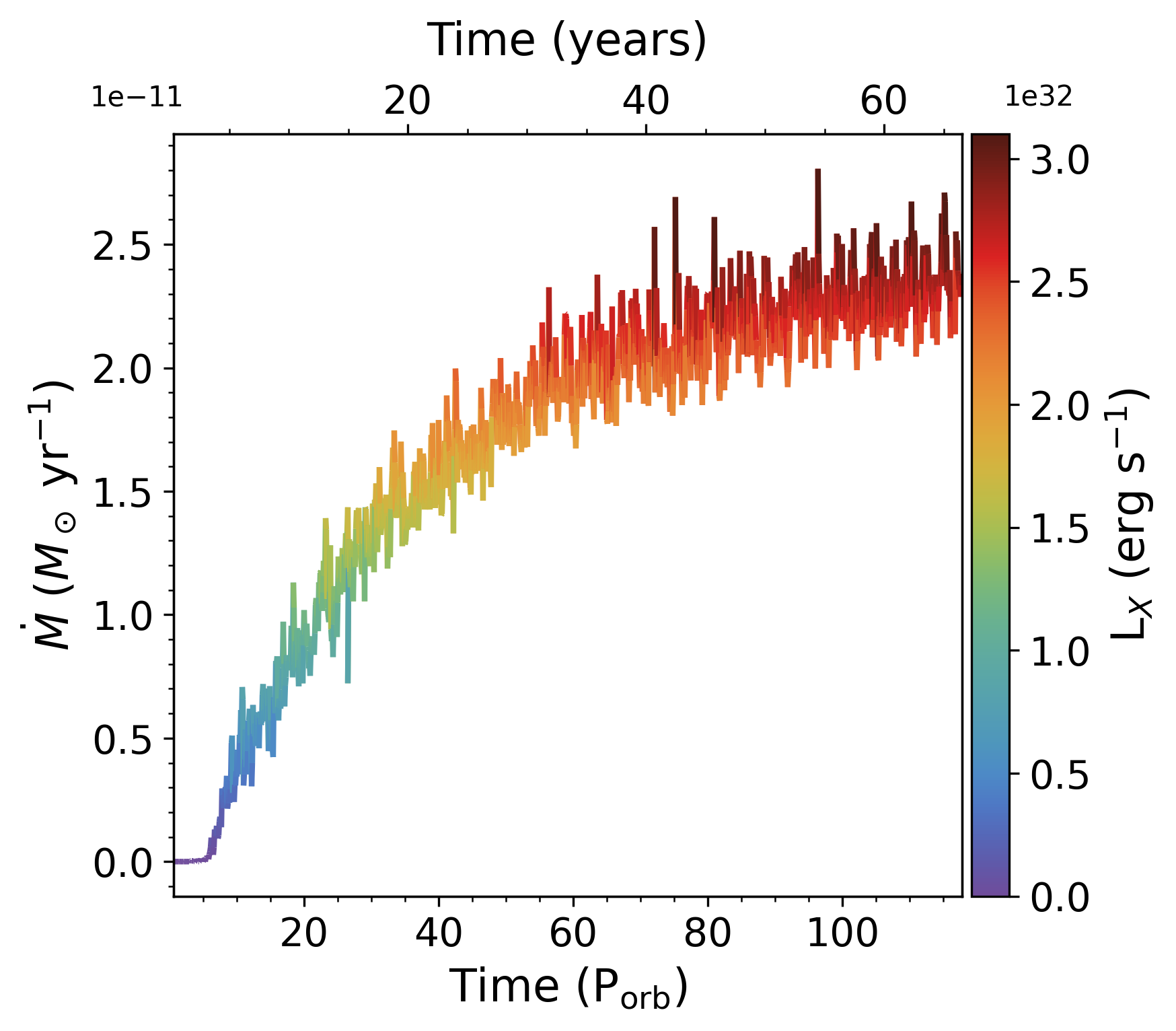}}\hfill
\subfloat{\includegraphics[width=\linewidth]{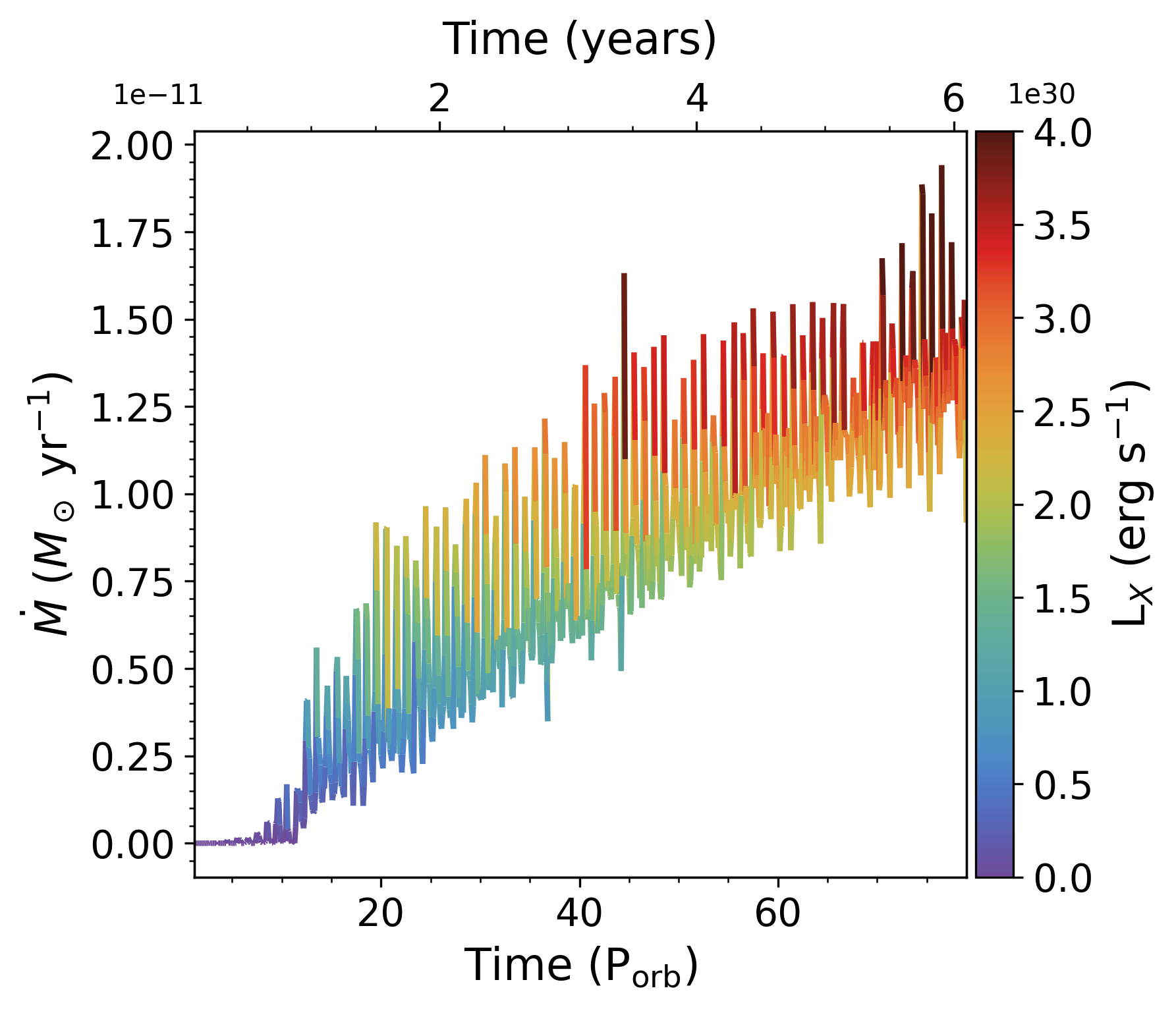}}\par 
\subfloat{\includegraphics[width=\linewidth]{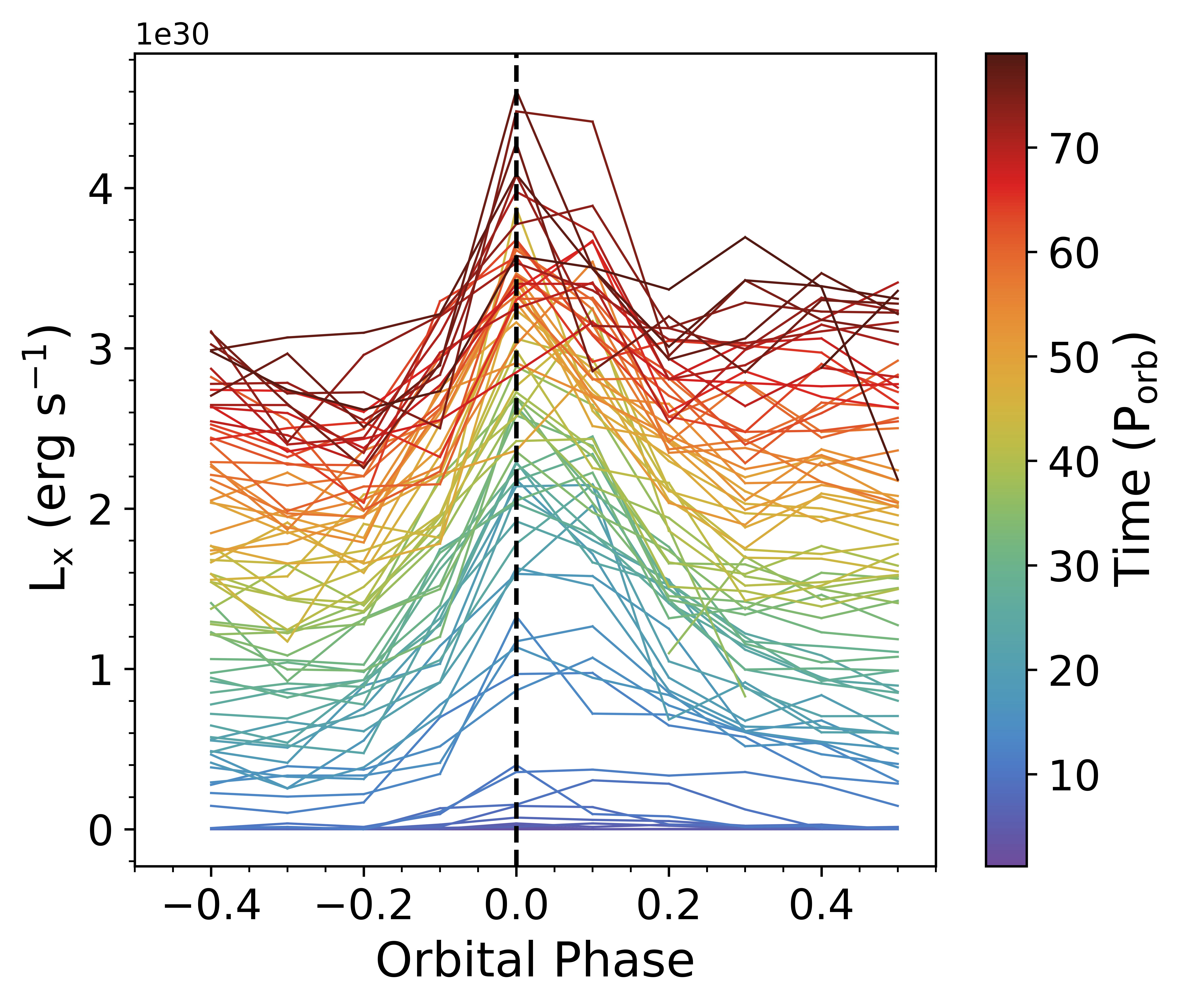}}
\caption{Accretion rates onto the secondary star as a function of time in the \textsc{sph} model for \gc\ (top) and 59\,Cyg (center). The colorbars show the predicted X-ray flux. Bottom: Predicted X-ray flux as a function of orbital phase for 59\,Cyg. The vertical dashed line represents the phase of periastron.}
\label{fig:mdot_lx}
\end{figure}

We can link our \textsc{sph} simulations to the systems' observed X-ray brightness by converting the measured accretion rates to a predicted flux. We estimated the X-ray flux that would be produced using
\begin{equation}
    L_X= \frac{\eta\, G M_X \dot{M}_X}{R_X}\,,
	\label{eq:xray_flux}
\end{equation}
where $M_X$ and $R_X$ are the mass and radius of the compact object, $\dot{M}_X$ is the accretion rate, and $\eta$ is the accretion efficiency, which we set to one as in \citet{oka01b} and \citet{rub25} to provide an upper bound for the luminosity. The top left panel of Fig.~\ref{fig:mdot_lx} shows the accretion rates and their corresponding X-ray luminosities for \gc, while the top right panel of Fig.~\ref{fig:mdot_lx} shows these values for 59\,Cyg. In both models, several orbital periods are completed before the disk extends far enough for particles to reach the secondary's accretion radius. Once this value is reached, the accretion rate continues to rise until the disk buildup period finishes. The length of time that this takes depends on the orbital period, the viscosity of the disk, and the mass ratio of the system. 

As the disk in the \gc\ model builds, the accretion rate is independent of orbital phase. On the other hand, the accretion rates seen in the simulation for 59\,Cyg are cyclic and phase-locked to the binary's orbital period. As we see in the bottom panel of Fig.~\ref{fig:mdot_lx}, accretion for 59$\,$Cyg is most efficient near and immediately following the periastron, and least efficient near apastron. These changes with orbital phase can be attributed to the non-zero eccentricity of this system. When the separation between the two stars is at a minimum, the secondary star is near a greater number of particles from the Be star disk than at other times in the orbit. As a result, more particles are captured by the secondary and the mass within the secondary’s Roche lobe is increased by $\sim$5\% compared to its minimum value. Variable accretion rates have been seen in \textsc{sph} simulations of Be/X-ray binaries by \citet{ras25} as well as asymptotic giant branch stars with eccentric, solar-mass companions, as found by \citet{mal24}.

In our models for both \gc\ and 59\,Cyg, the secondary star induces a $m = 2$ spiral density pattern in the Be star disk. Such density enhancements have been noted many times in \textsc{sph} simulations of Be binary systems with various orbital parameters (e.g., \citealt{oka02, pan16, cyr17}) and were studied in detail in \citet{cyr20}. Observations of \gc\, in particular, have showed the presence of these $m = 2$ waves. \cite{bor20} and \citet{baa23} noted phase-locked variability in the emission profile of 59 Cyg which may also be consistent with the same structure. In our simulations, the spiral arms dominate the Be star disk between the dense, inner region and the truncation radius. In the \gc\ simulation, the leading spiral arm forms a structure joining the Be disk and the Roche lobe of the secondary star, referred to as the ``bridge" by \citet{rub25}. This bridge is maintained throughout an orbital period. It provides a channel for matter to cross the Roche equipotential near the L1 point and feed an accretion disk around the secondary star. This is evident in the top panels of Fig.~\ref{fig:splash}. The secondary's distance from the Be star disk does not vary with time in our circular model for \gc. As a result, the size of the secondary's Roche lobe stays constant and the density of the bridge is independent of orbital phase. The rate at which matter enters the secondary's Roche lobe in the \gc\ simulation, then, is therefore sustained throughout an orbital period. The predicted luminosity, which is directly proportional to the accretion rate, is also constant with orbital phase. 

Structural changes to the bridge also play a role in the phase-dependent accretion rate and luminosity seen in 59\,Cyg. The Roche lobe of the secondary star in this slightly eccentric system varies with orbital phase according to Equation~\ref{eq:roche_lobe}. Additionally, the bridge of connecting material only exists for some orbital phases. As we see in the bottom panels of Fig.~\ref{fig:splash}, the bridge forms near periastron, feeds the secondary's accretion disk for a short time afterward, and then disappears prior to apastron. 

Most \gc\ objects are known to have X-ray luminosities of $\sim$\SI{e31}-\SI{e33} ergs s$^{-1}$ \citep{ste12, naz20, rau22b}. \gc\ itself has been observed to have an X-ray luminosity of $\sim$\SI{e32}-\SI{e33} ergs s$^{-1}$ \citep{par93, smi16}. Our predicted flux of $\sim$\SI{2e32}{} ergs s$^{-1}$ is therefore consistent with the lower limit of the observations for \gc. The predicted X-ray luminosities for 59\,Cyg, whose sdO companion has a larger radius than the WD we modelled for \gc, are $\sim$\SI{3e30}{} ergs s$^{-1}$, well below the threshold expected for a \gc\ analog. This is in agreement with observations of this system, which is known to be a soft and faint X-ray source lacking the luminosity and hardness characteristic of \gc\ objects \citep{naz20, naz22c}.

Several factors contribute to the difference in the predicted X-ray luminosities for these systems. In the simulation for 59\,Cyg, the transition radius is smaller than for \gc\ since its orbital period is shorter. As a result, the disk in the 59\,Cyg model takes fewer orbital periods to reach its maximum radial extent, and can achieve relatively high densities within that time. However, while the disk for \gc\ takes more orbital periods to grow to its full radius, it has more time during each orbital period to accumulate mass. After building for the same number of orbital periods, \gc\ has built a more massive disk; for example, after 50 P$_{\rm{orb}}$, its total disk mass is $\sim$\SI{1.5e-9}{} M$_{\odot}$, more than double the value for 59\,Cyg. The rates at which particles enter the secondary's accretion radius, then, will be very different between the two models. Since the predicted X-ray luminosity in Equation~\ref{eq:xray_flux} is a function of this accretion rate, it is expected that the X-ray luminosity produced by the 59\,Cyg simulation after the same number of orbital periods is significantly smaller than for \gc. Additionally, the nature of the companion star has a significant impact on the predicted X-ray flux because the luminosities are dependent on the mass and radius values chosen for the accreting star. The secondary's significantly smaller radius and somewhat larger mass in \gc\ compared to 59\,Cyg also plays a role in its larger predicted X-ray luminosities after the same number of periods. By Equation \ref{eq:xray_flux}, we would obtain X-ray luminosities for 59\,Cyg that are $\sim$40 times larger than the given values if we assumed the companion was a WD rather than an sdO. We also note that the size of the gravitational well around the secondary star in these systems differs due to their distinct masses, which impacts the hardness of the observed X-ray flux.

The accretion process in our models bears resemblance to the Roche lobe overflow commonly seen in close binaries, but it has important differences. Since the Be star's decretion disk is rotating, and the secondary causes significant tidal friction in the disk, mass transfer does not occur exactly at the L1 point. Our models also show key departures from the standard implementation of Bondi–Hoyle–Lyttleton (BHL) accretion trends \citep{hoy39, bon44}, which are often applied to binary systems where Roche lobe overflow is absent and accretion is driven by winds \citep{edg04}. Importantly, the BHL accretion model is only applicable to systems where the wind velocity ($v_w$) is much larger than the orbital velocity of the accretor \citep{bof15, han16}. The Hoyle-Lyttleton accretion rate is given by $\dot{M}_{\rm{HL}} = \pi \zeta_{\rm{HL}}^2 v_w \rho$ where $\zeta_{\rm{HL}}=2GM/v_w^2$ is the impact parameter with $M$ as the mass of the accreting body, $\rho$ being the density of the accreting material and $v_w$ as the relative velocity of the flow \citep{edg04}. Since each secondary star was modelled as coplanar with the Be star disk, and these disks are quasi-Keplerian \citep{wat00, krt11}, we can approximate $v_w$ as the radial velocity of the disk with respect to the secondary star. In our simulations for both \gc\ and 59\,Cyg, the orbital velocities of the secondary stars are comparable to (or slightly exceeding) $v_w$ using this definition. Therefore, applying the BHL model without correction yields nonphysical estimates that are inconsistent with our predicted values in Figure~\ref{fig:mdot_lx}. However, \citet{tej25} recently presented improved analytical models for cases where the wind
speed is comparable to the orbital speed, by applying a geometric correction. Using their circular model for \gc, we find rates of $\approx$\SI{1.4e-11}M$_{\rm{\odot}}$/yr, comparable to our predicted values. Similarly, applying their approximations for elliptical orbits to 59\,Cyg yields $\approx$\SI{e-10} M$_{\rm{\odot}}$/yr; while larger than our predicted values, it provides a better match than an uncorrected BHL model by orders of magnitude. 

It is interesting to consider whether the simulated accretion of material by the secondary stars in these systems could eventually produce Type Ia supernovae or classical novae. The accretion rate for \gc\ stabilizes near \SI{2e-11}{} M$_{\odot}$ yr$^{-1}$, while the model for 59\,Cyg has maximum values closer to \SI{1e-11}{} M$_{\odot}$ yr$^{-1}$. With the masses of the secondary stars listed in Table~\ref{tab:sys_params}, it would take on the order of $10^{10}$ years for either of these systems to reach the Chandrasekhar limit. Given the main sequence lifetimes of B-type stars, and the fact that the Be phenomenon is restricted to main sequence or slightly evolved stars, it is not likely that the mass transfer from the disk could be sustained long enough to produce supernovae in the future. Nova recurrence rates for \gc\ can also be estimated from WD accretion models; using those of \citet{wol13} and \citet{cho21}, our predicted values correspond to nova recurrence timescales larger than \SI{e7} years. Other works have predicted companion accretion rates on the order of \SI{e-10}{} M$_{\odot}$ yr$^{-1}$ for \gc\ \citep{tsu18, gun25, toa25}, and nova recurrence times for such a 1 M$_{\odot}$ WD would exceed \SI{e5} years \citep{gun25, toa25}. In light of these estimates, the utter lack of Type Ia supernova and nova detections among \gc\ analogs to date seems reasonable.

\begin{figure}[!tbp]
  \begin{subfigure}[b]{\linewidth}
    \includegraphics[width=\linewidth]{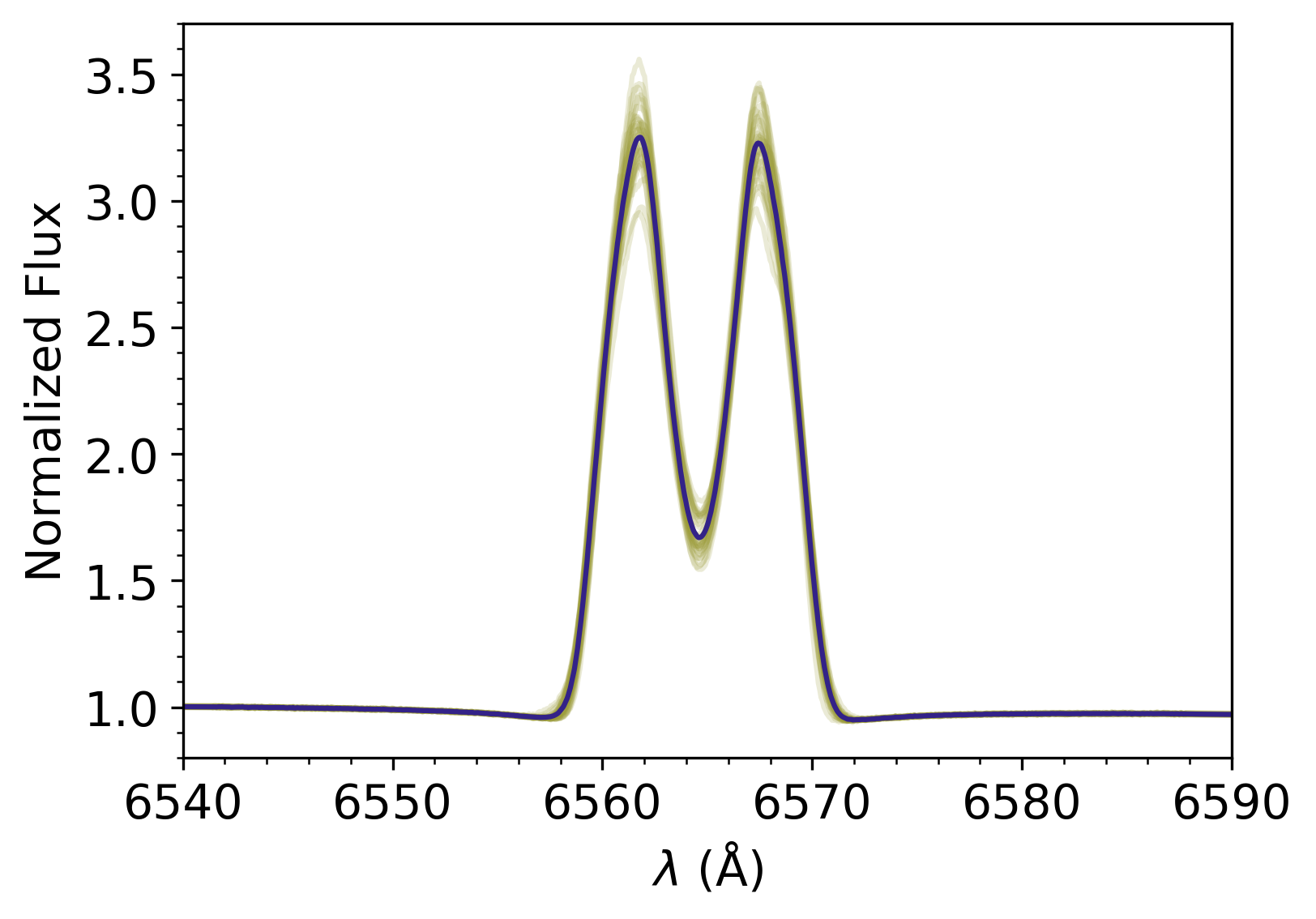}
  \end{subfigure}
  \hfill
  \begin{subfigure}[b]{\linewidth}
    \includegraphics[width=\linewidth]{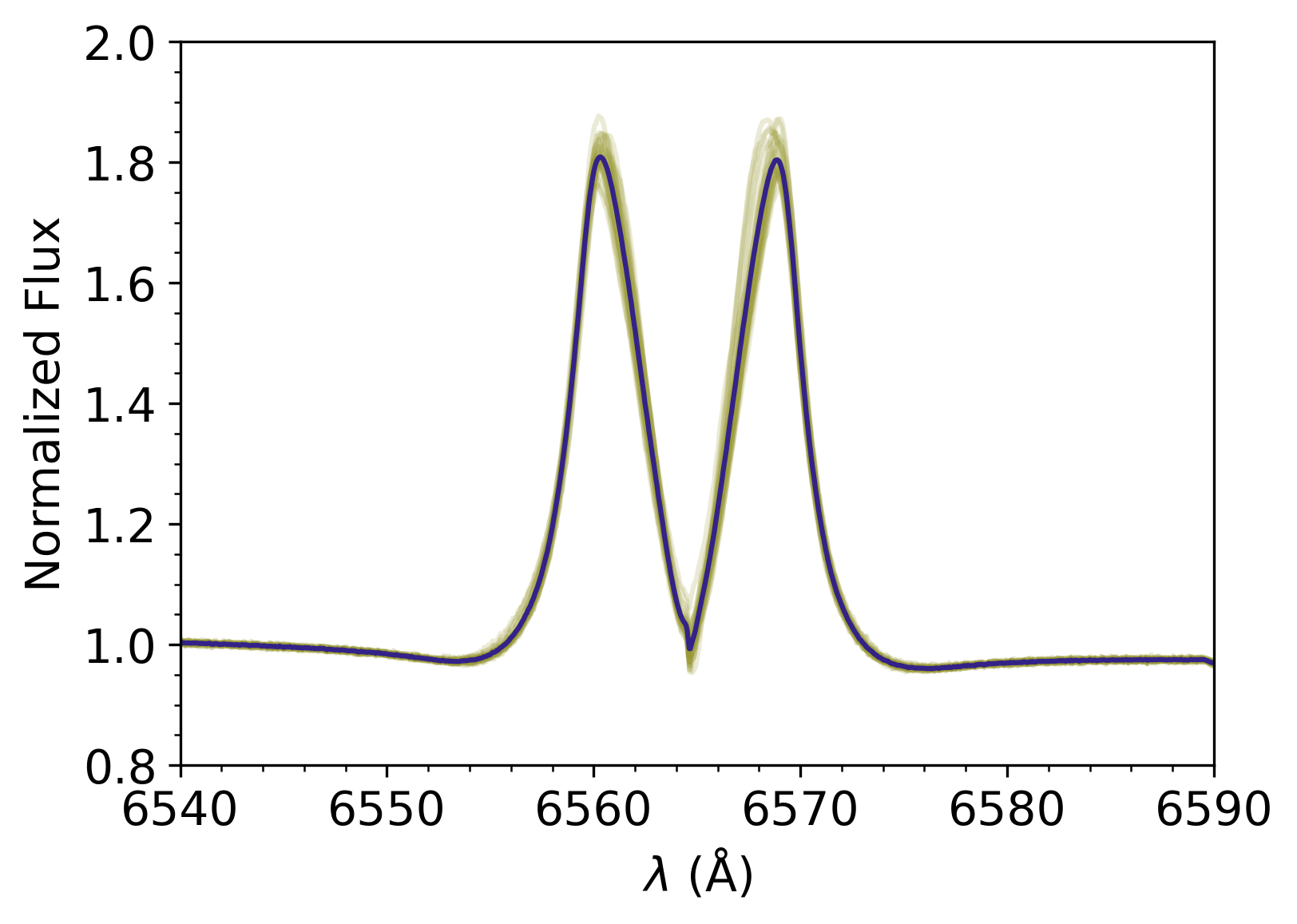}
  \end{subfigure}
  \caption{
  Predicted  H$\alpha$ line for \gc\ (top) and 59\,Cyg (bottom). The dark line indicates the flux values averaged over all simulations, while individual simulations are shown in lighter green. For both systems, the profile is shown at an azimuthal angle of $\phi=0$.}
  \label{fig:ha_lines}
\end{figure}

\begin{figure*}[!tbp]
  \begin{subfigure}[b]{0.32\textwidth}
    \includegraphics[width=\textwidth]{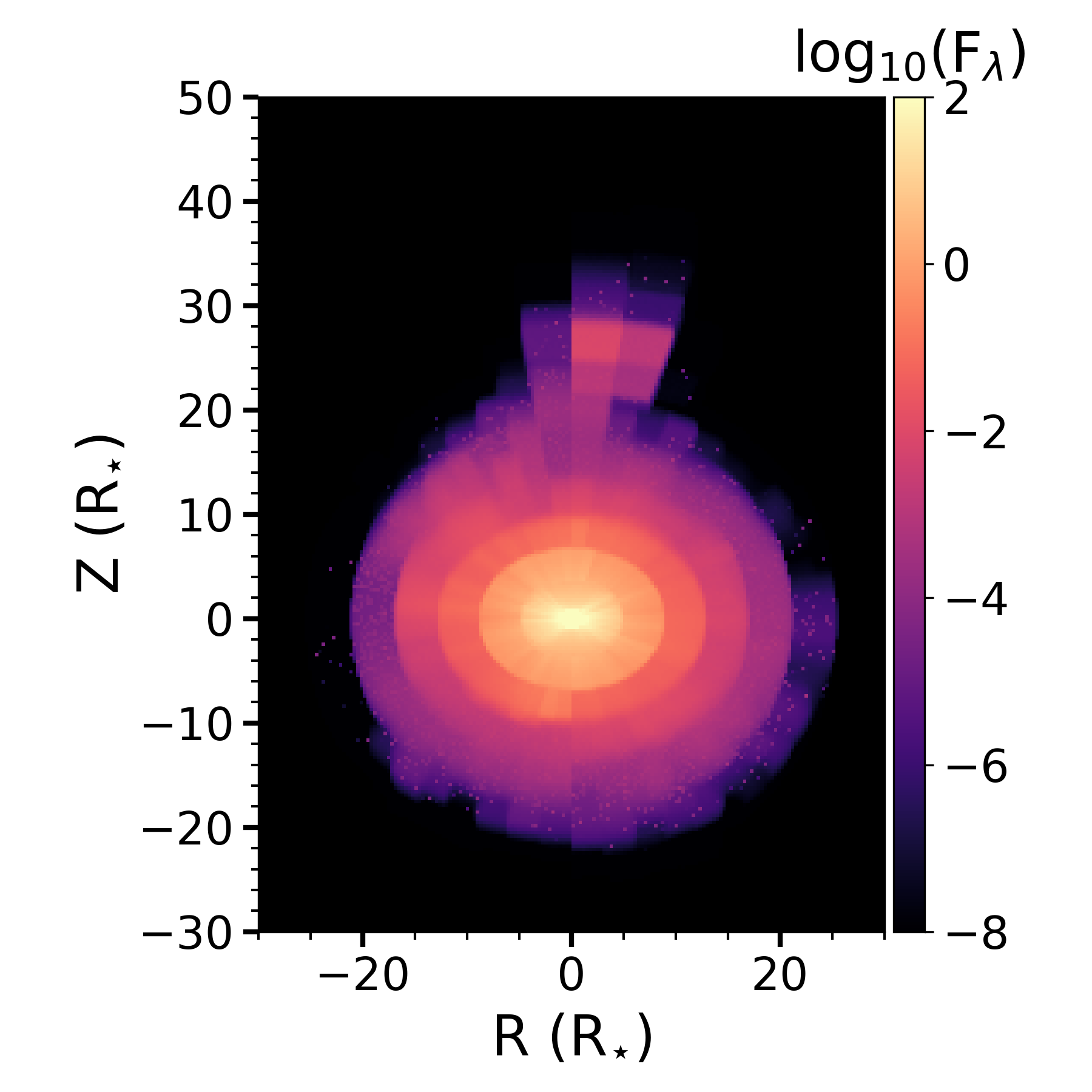}
  \end{subfigure}
  \begin{subfigure}[b]{0.32\textwidth}
    \includegraphics[width=\textwidth]{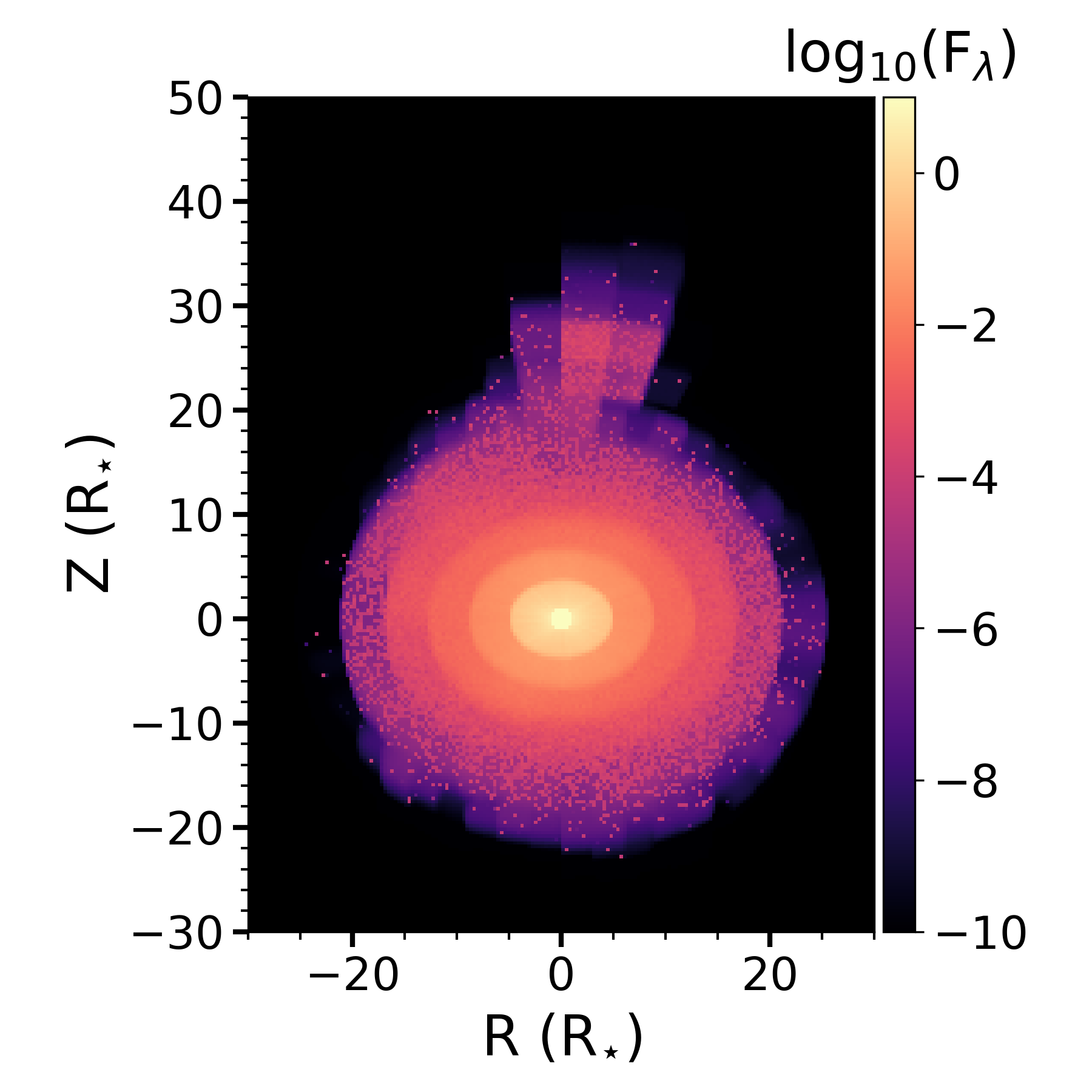}
  \end{subfigure}
  \begin{subfigure}[b]{0.32\textwidth}
    \includegraphics[width=\textwidth]{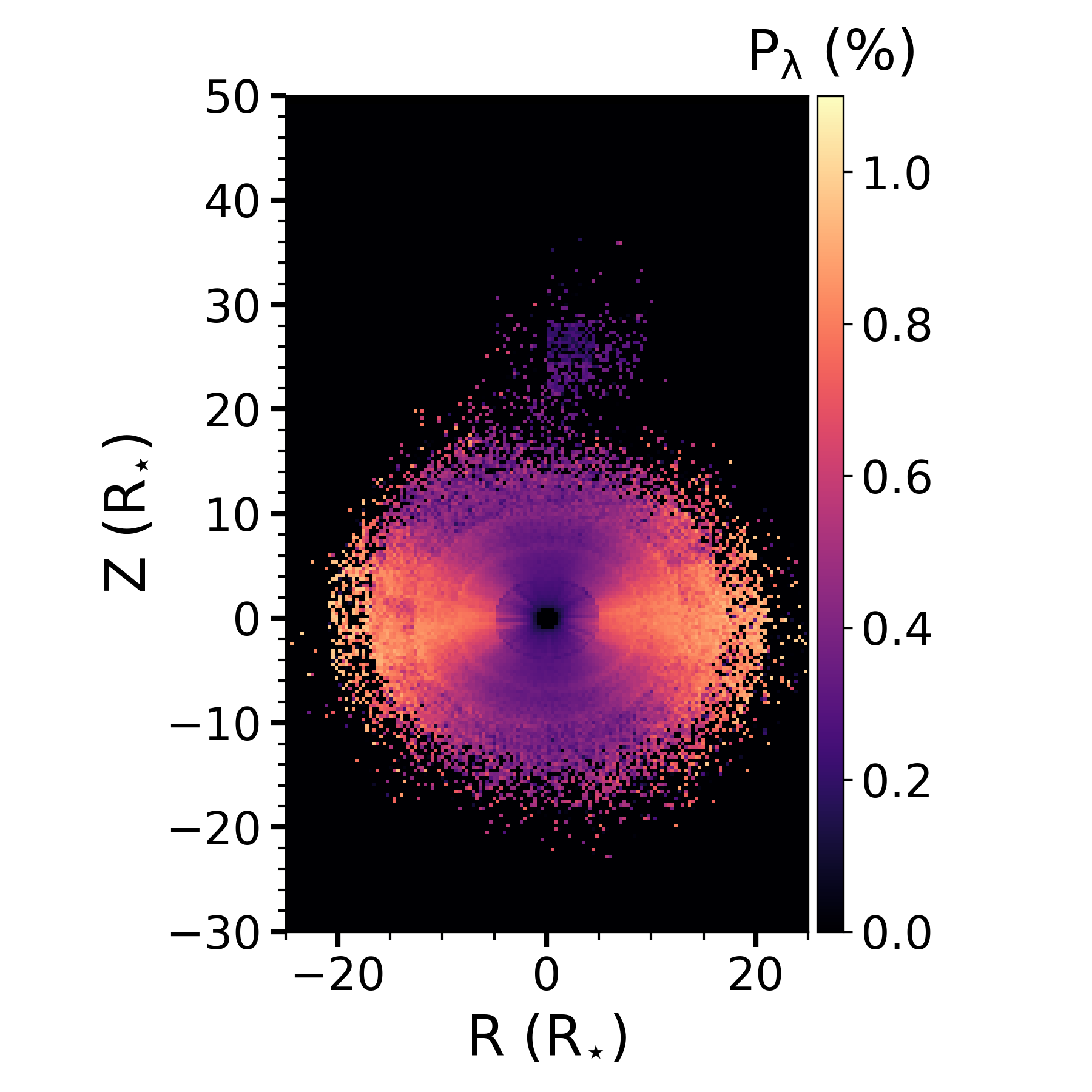}
  \end{subfigure}
  \begin{subfigure}[b]{0.32\textwidth}
    \includegraphics[width=\textwidth]{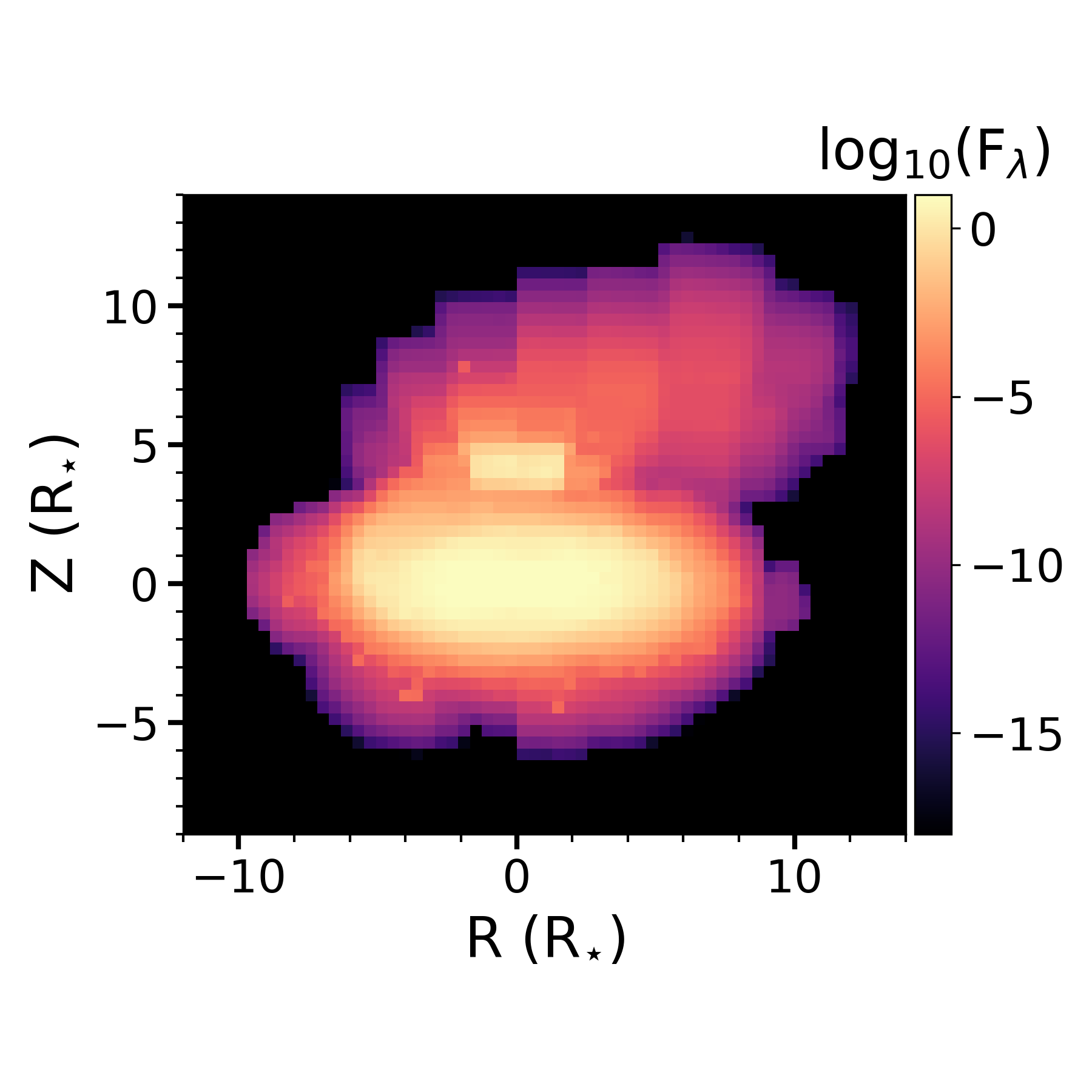}
  \end{subfigure}
  \begin{subfigure}[b]{0.32\textwidth}
    \includegraphics[width=\textwidth]{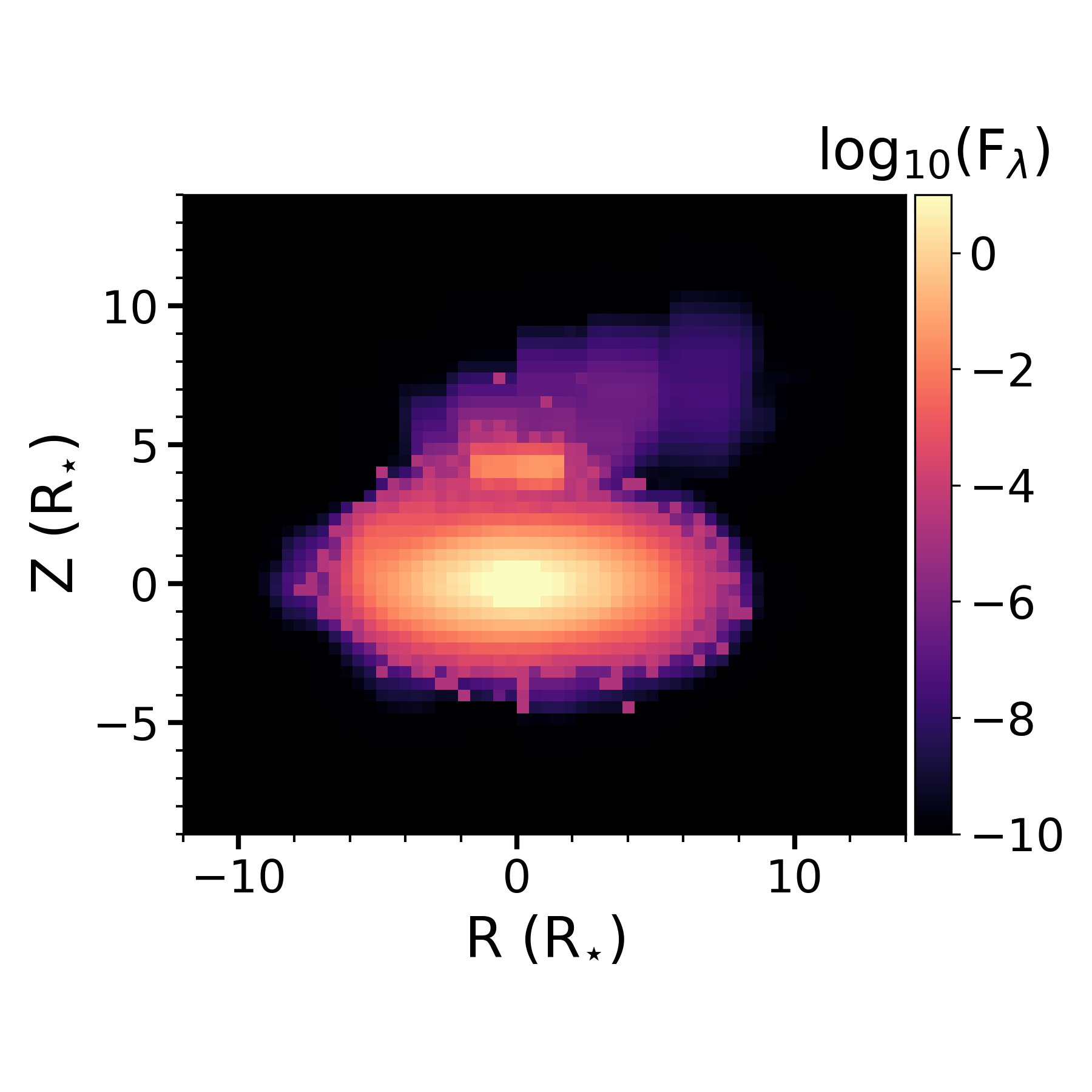}
  \end{subfigure}
  \begin{subfigure}[b]{0.32\textwidth}
    \includegraphics[width=\textwidth]{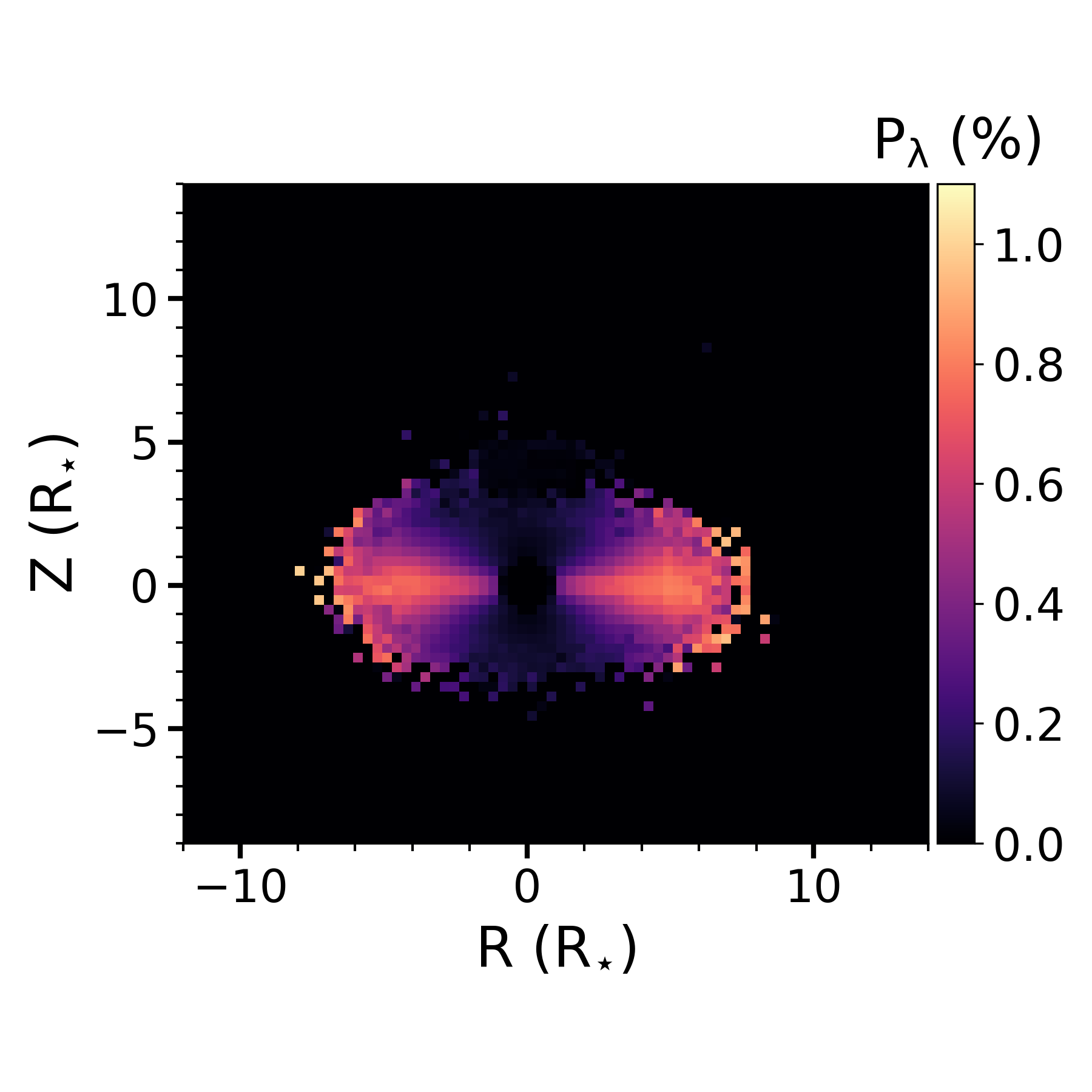}
  \end{subfigure}
  \caption{The disk image computed by \textsc{hdust} for \gc\ at an inclination angle of 43$^{\circ}$ (top), and 59\,Cyg at an inclination angle of 70$^{\circ}$ (bottom), colored by flux emitted at H$\alpha$ (left) and UV (center) wavelengths, as well as the polarization degree across UV wavelengths (1000-3500 \AA, right).}
  \label{fig:disk_images}
\end{figure*}

\begin{figure}[!tbp]
  \begin{subfigure}[b]{\linewidth}
    \includegraphics[width=\linewidth]{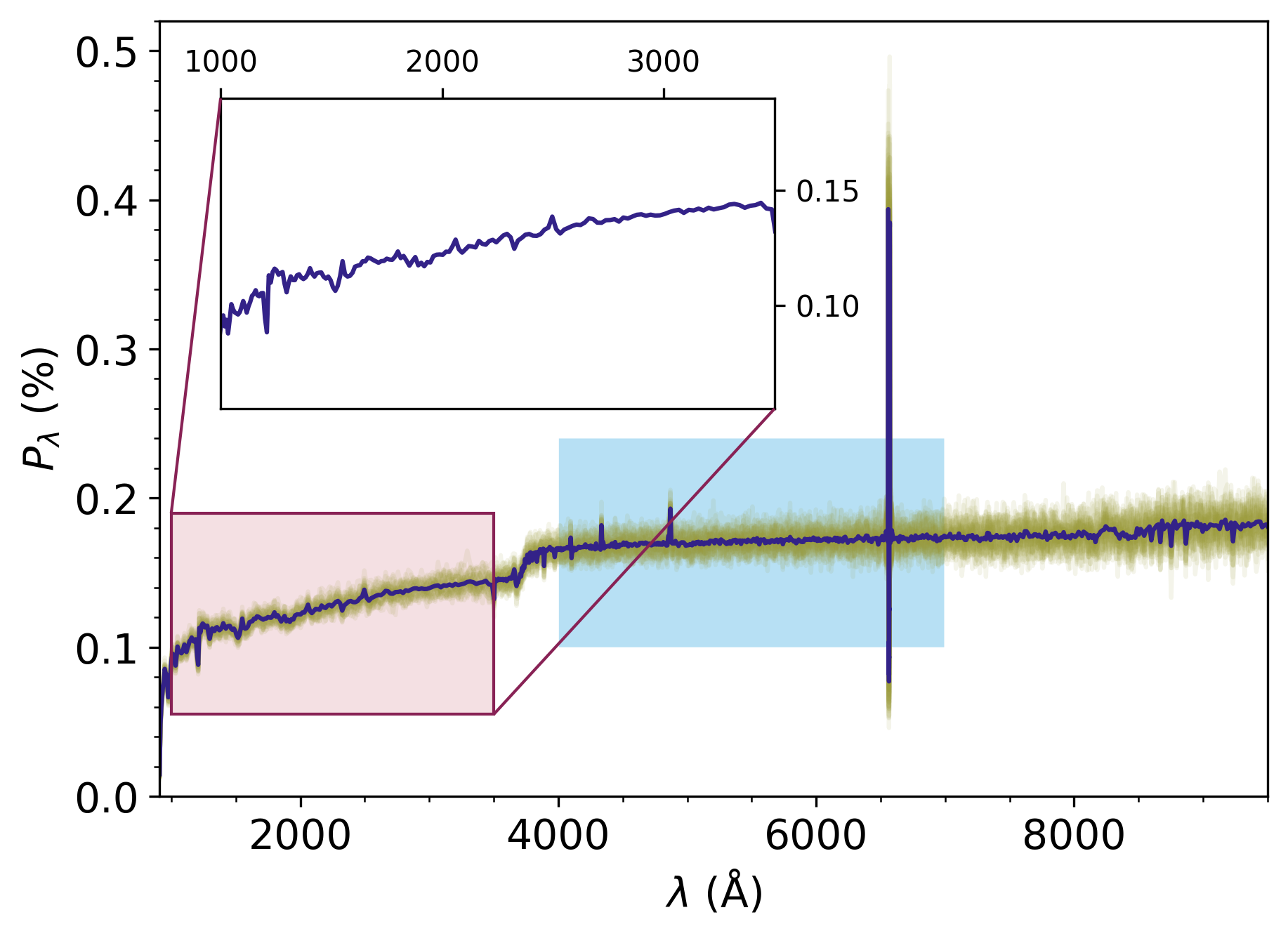}
  \end{subfigure}
  \hfill
  \begin{subfigure}[b]{\linewidth}
    \includegraphics[width=\linewidth]{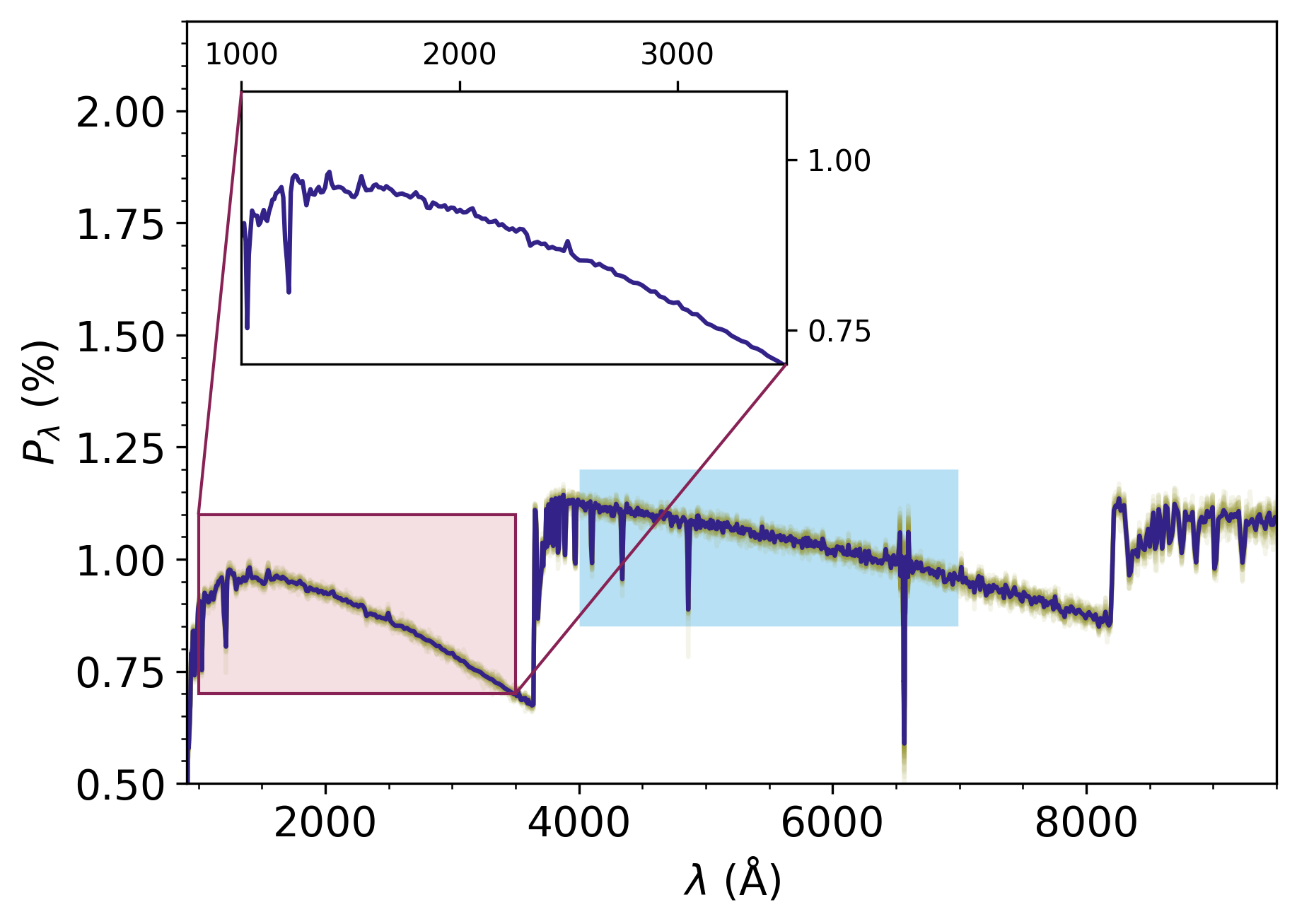}
  \end{subfigure}
  \caption{Predicted polarized spectrum for \gc\  (top) and 59\,Cyg (bottom). The dark line indicates the polarized spectrum averaged over all simulations, while individual simulations are shown in lighter green. The wavelengths used to calculate the average polarization across the UV (1000-3500 \AA) and V (4000-7000 \AA) bands are shaded in the left and right shaded regions, respectively. An enlarged view of the UV region is provided in the inset.}
  \label{fig:pol_spectrum}
\end{figure}

We averaged the H$\alpha$ profiles produced by the thirty \textsc{hdust} simulations for both \gc\ and 59\,Cyg, and present the averaged profiles in Fig.~\ref{fig:ha_lines}. Our models did not indicate phase-dependent behaviour, but a higher time resolution would be required to detect these variations and was beyond the scope of this work. As expected, the H$\alpha$ emission is more prominent in \gc, due to the larger extent of the H$\alpha$ emitting area. The H$\alpha$ line is generally estimated to be formed in the innermost $\sim10-20$ $R_{\star}$ of the Be disk \citep{car11}, so the density and structure of the disk within this radius determine the strength and morphology of the H$\alpha$ line. The top left panel of Fig.~\ref{fig:disk_images} shows that the bulk of the H$\alpha$ emitting region is indeed found within the first 10-15 $R_{\star}$, with some extending to roughly 20 $R_{\star}$. By comparison, the H$\alpha$ emitting region in 59\,Cyg, shown in the bottom left panel of Fig.~\ref{fig:disk_images}, shows that the bulk of the H$\alpha$ flux originates in the innermost $\sim$7 $R_{\star}$, as it has been truncated by the smaller orbital separation of the binary. When comparing the two systems at the same number of orbital periods, \gc\ has had more time (in years) to build its disk, and is not truncated by the shorter orbital period, so it has a larger H$\alpha$ emitting area, as expected.

Since its discovery, \gc\ has been observed both with and without a disk \citep{bal40, nem12}. However, its Balmer lines have been consistently in emission since the late 1940s. The relative fluxes of our normalized H$\alpha$ lines are consistent with those reported by \citet{nem12} and \citet{bor20} (see also the many H$\alpha$ observations available in the BeSS database\footnote{\url{http://basebe.obspm.fr/}}). The peak intensities of our profiles for 59\,Cyg are also consistent with the archival profiles presented in \citet{baa23} and available on the BeSS database. The profile morphologies of our models do not replicate the flat-topped profiles often seen in 59 Cyg (see e.g. \citealt{baa23}). However, exactly reproducing line profiles is a process with several degeneracies as different combinations of density, rate of density decrease, and inclination can produce similar line profiles. It has been suggested that both systems undergo changes in disk inclination \citep{baa23} which further complicates such an endeavor. Importantly, our \textsc{sph} models are able to produce H$\alpha$ emission that is comparable to measured values. We leave further detailed modelling efforts for future work. 

As with the H$\alpha$ profiles, we estimated the intrinsic polarization values expected from these systems by averaging the polarization degrees from the \textsc{hdust} simulations over five orbital periods. The intrinsic polarization predicted by \textsc{hdust} is produced by Thomson scattering in the disk, which is  modelled as devoid of dust. We computed the average polarization over the V (4000-7000 \AA) band and found values of 0.18\% and 0.96\% for \gc\ and 59 Cyg, respectively. At UV wavelengths (1000-3500 \AA), we estimate the polarization degree for \gc\ at 0.11\% and 59\,Cyg at 0.95\%. As demonstrated by the polarized spectra shown in Fig.~\ref{fig:pol_spectrum}, the polarization signal in the 59\,Cyg model is much larger than that of \gc\ due to the enhanced density in its truncated disk.

Historical polarization degree measurements of \gc\ using the University of Winsconsin's HPOL spectropolarimeter have reported V-band values near 0.4\%-0.5\% from 1990 to 2005 \citep{dra14}. Using the University of Ontario's photoelectric Pockels cell polarimeter, \citet{poe77} found intrinsic polarization values of $\sim$0.3\%-0.5\% for the H$\alpha$ line from 1974-1976. These estimates are substantially larger than our estimate of  0.18\% across the V-band. We note that our predicted values are consistent with the low-density disk in our \textsc{sph} model, which had a base density on the order of \SI{e-12}{\gram \per \cubic \centi \meter}. As discussed below, disk density has a strong impact on the observed polarization degree and shape of the polarized continuum. Varying this parameter in our \textsc{hdust} simulations is expected to significantly affect the predicted polarization values, but a detailed modelling effort on \gc\ over time was not the focus of this work.

The polarization degree for 59\,Cyg has varied over time, with measurements in the 1970s reporting between 0.03\% at 8490 \AA\ and 0.83\% over 6510-6525 \AA\ \citep{poe76, poe79}. The V-band polarization, uncorrected for interstellar polarization, was reported as 0.57\% by \citet{hua89} using data from the MacDonald Observatory's Breger polarimeter; it was also monitored from 1992-1994 and found to be 0.33-0.42\% \citep{dra14}, staying roughly constant at 0.32-0.44\% from 1995 to 2004 \citep{dra14}. Our average V-band polarization value of 0.96\% is not far from the value reported by \citet{poe76} for 6510-6525 \AA\ but is larger than the V-band values measured in the 1990s to early 2000s. Polarization at UV wavelengths has not been consistently measured, with the best comparison being 0.35\% at 3440 \AA\ in 1975 \citep{poe79}, which is significantly smaller than our estimate. As above, we note that a dedicated effort to replicate the observed polarization values using \textsc{hdust} models was beyond the scope of this work. Notably, our models correctly predict higher intrinsic polarization for 59\,Cyg than in \gc.

The shapes of the polarized spectra in Fig.~\ref{fig:pol_spectrum} are strongly dependent on the density of the disk that produces the signal. The density in the innermost region in the disk is on the order of \SI{e-12}{\gram \per \cubic \centi \meter} for \gc\ and \SI{e-11}{\gram \per \cubic \centi \meter} for 59\,Cyg. The truncated, denser disk for 59\,Cyg therefore produces a stronger polarization signal. In low density regimes, the disk opacity is dominated by the wavelength-independent electron scattering opacity \citep{hau14}. This explains why the polarized spectrum for \gc, which has a relatively lower density, is nearly flat. Meanwhile, for larger densities such as those seen in 59\,Cyg, the bound-free and free-free opacities become the dominant opacity sources in the disk, since they scale approximately with the square of the density while the electron opacity increases roughly linearly \citep{bjo94, hau14}. The increased contribution from bound-free and free-free opacity results in steeper continuum slopes. Additionally, the increase in H\,\textsc{i} opacity for denser disks produces a more pronounced Balmer discontinuity \citep{woo96}. The enhanced polarization seen in 59\,Cyg is consistent with the predictions of \citet{rub25}, who suggested that the ``accumulation effect", where the disk density is elevated by the accumulation of matter within the transition radius, would enhance the polarization levels from Be binary systems.  

We note that \textsc{hdust} does not currently include the effects of UV line blanketing, since metals are not included in the code. Line blanketing, especially in iron lines, is expected to cause a reduction in the polarization levels at UV wavelengths. Therefore, the polarization degree we report here represents the upper limit for the tested disk densities. 

Overall, we are able to produce X-ray luminosities and H$\alpha$ profiles that are consistent with observations for both \gc\ and 59\,Cyg. An accreting compact object can account for the large X-ray fluxes observed in \gc. The faint, soft X-rays observed in 59\,Cyg are consistent with an sdO companion. The polarization levels that are produced by these systems are detectable at both visual and UV wavelengths, making them compelling objects for future study in the UV.

\subsection{Future UV monitoring of \gc\ analogs} 
\label{subsec:polstar_links}

UV observations are expected to reveal important information on the source of the X-ray production mechanism in \gc\ analogs. Up to now, the observed UV continuum emission in \gc\ has shown variation over timescales of hours, and spectral features associated with the wind have also changed with time. However, it is difficult to ascertain how typical this is of Be stars in general, or unique to \gc\ analogs, due to how few objects have been thoroughly studied in this range. Large-scale monitoring at UV wavelengths of \gc\ analogs, and other Be stars for comparison, is thus crucial for resolving the similarities and differences in these systems. With the opportunities newly available from the instrument and mission concept designs such as those proposed for {\em Polstar}, a proposed Small Mission Explorer concept for UV spectropolarimetry designed to study the influence of rapid rotation in massive stars \citep{paulNew}, {\em Pollux}, a proposed UV, visible and IR very high resolution spectropolarimeter for the future NASA flagship HWO, and {\em Arago}, a UV and visible spectropolarimeter mounted on a 1-m telescope proposed to ESA \citep{mus23}, we expect systems like \gc, 59~Cyg, and MV Lyrae will be observable in the UV regime within the next few decades.

UV spectroscopy has been a cornerstone of astronomical inquiry for the last several decades \citep[e.g.,][]{Linsky2018}. The success of missions such as the International Ultraviolet Explorer (1978-1996) and the Hubble Space Telescope (1990-present) have revealed the UV universe in stunning detail. Now, as the James Webb Space Telescope\footnote{\url{https://science.nasa.gov/mission/webb/}} and anticipated Nancy Grace Roman Space Telescope\footnote{\url{https://science.nasa.gov/mission/roman-space-telescope/}} open new windows into the formation of early galaxies and cycles of stellar feedback \citep[e.g.,][]{atek2024}, the need to advance complementary UV capability is clear.

In particular, the potential contribution of modern UV spectropolarimetry has recently been highlighted by several pilot studies, such as those by \citet{Jones2022}, \citet{Peters2022}, and \citet{udDoula2022}. Spectropolarimetry provides a powerful diagnostic for the complex, asymmetric circumstellar environments of rapidly rotating early-type stars like \gc\ \citep[e.g.,][]{Ignace2025, ras25b}. UV radiation is sensitive to hot material, and the polarized spectrum is sensitive to the density of Be star disks. Thus, UV spectropolarimetric studies are expected to enable an improved characterization of Be star disks.

\section{Discussion and conclusions}
\label{sec:discussion_conclusions}

The origin behind the hard X-rays emitted by \gc\ analogs continues to be elusive, hindered by the absence of a dedicated multiwavelength monitoring campaign on these systems. This paper has focused on the possibility that the observed X-ray flux could be produced by an accreting WD companion, in contrast with other Be binary systems with sdO companions. We investigated two scenarios using \textsc{sph} simulations: one model for \gc\ with a WD companion, and another for 59\,Cyg, a non-\gc\ Be star, known to have an sdO companion. We find that the secondary star is able to siphon matter from the Be star decretion disk to form its own disk-like structure in both systems, similar to the accretion disks seen in \textsc{sph} simulations of Be/X-ray binaries with accreting neutron star companions \citep{mar14, fra21, ras25, rub25}. These disks seem to have Keplerian rotation velocities as well as scale heights that roughly follow theoretical predictions within the inner regions. The structure around the secondary star in the \gc\ system roughly follows the density distribution expected for an accretion disk, while the circumsecondary of 59\,Cyg has a slightly more complex density profile. 

Our simulations show that infalling material could generate X-ray fluxes on the order of \SI{e32}{\erg \per \second} for \gc, on the lower end of the observed values for this system, and \SI{e30}{\erg \per \second} for 59\,Cyg, which is consistent with a faint X-ray source. The lower X-ray luminosity of 59\,Cyg compared to \gc\ can be attributed to two factors: the larger radius of the sdO companion compared to the WD in \gc, and the rate at which matter is accreted by the secondary, which is largely dependent on the mass ratio (and therefore the size of the Roche lobe) of the primary and secondary. The slightly eccentric orbit of 59\,Cyg also results in phase-dependent accretion rates and resulting X-ray fluxes. 

The predicted polarization degree for both \gc\ and 59\,Cyg is somewhat smaller across UV wavelengths compared to visual wavelengths, but still potentially detectable. The polarization degree in 59\,Cyg is significantly larger than in \gc, largely due to its smaller orbital period and denser, truncated disk resulting in a larger number of scatterers. The polarization's roughly linear sensitivity to the disk density is complementary to optical emission processes that are sensitive to density squared. With the advent of new UV spectropolarimetric facilities \citep{nei25}, it may soon be possible to produce detailed characterizations of \gc \, and its analogs in both spectral and temporal domains. Fig.~\ref{fig:pol_spectrum} above indicates that the polarized UV spectrum of \gc \, is expected to be observable at a level of $p_{\lambda} \sim 0.1 \%$, and with that of 59~Cyg at least a factor of eight higher. The current proposed designs of {\em Polstar} \citep{paulNew}, {\em Pollux} \citep{Pollux2024}, and {\em Arago} \citep{mus23} should achieve polarization sensitivities of at least $p_{\lambda} \sim 0.1 \%$, making UV polarimetric studies of objects such as \gc{} and 59~Cyg achievable for the first time.    

\backmatter

\bmhead{Acknowledgments}
The authors thank the referee, Jes\'us A. Toal\'a, for their helpful comments and suggestions which improved the paper. We also thank AC Carciofi's research group and Atsuo Okazaki for insightful discussions. YN, a senior research associate from FNRS, acknowledges support from FNRS and the Li\`ege university. RGR and CEJ acknowledge support from the Natural Sciences and Engineering Research Council of Canada. AuD acknowledges support from NASA through Chandra Award number TM4-25001A issued by the Chandra X-ray Observatory 27 Center, which is operated by the Smithsonian Astrophysical Observatory for and on behalf of NASA under contract NAS8-03060. JLB is co-funded by the European Union (ERC, MAGNIFY, Project 101126182). Views and opinions expressed are, however, those of the authors only and do not necessarily reflect those of the European Union or the European Research Council. Neither the European Union nor the granting authority can be held responsible for them. This research has made use of the SIMBAD database operated at CDS, Strasbourg (France), and of NASA’s Astrophysics Data System (ADS). This work was made possible through the use of the Shared Hierarchical Academic Research Computing Network (SHARCNET). We acknowledge the use of \textsc{splash} \citep{pri07} for rendering and visualization of our figures.

\section*{ORCID iDs}

RGR \url{https://orcid.org/0009-0007-9595-2133} \\
YN \url{https://orcid.org/0000-0003-4071-9346} \\
JLB \url{https://orcid.org/0000-0002-2919-6786} \\
CEJ \url{https://orcid.org/0000-0001-9900-1000} \\
CE \url{https://orcid.org/0000-0003-1299-8878} \\
KG \url{https://orcid.org/0000-0001-8742-417X} \\
AuD \url{https://orcid.org/0000-0001-7721-6713} \\
CN \url{https://orcid.org/0000-0003-1978-9809} \\
JD \url{https://orcid.org/0000-0002-0210-2276}


\bibliography{gcas_ref}

\section*{Declarations}
RGR received funding from the Natural Sciences and Engineering Research Council of Canada (NSERC) Postgraduate Scholarship - Doctoral program. CEJ received support from the NSERC Discovery Research program. AuD has received research support from NASA through Chandra Award number TM4-25001A issued by the Chandra X-ray Observatory 27 Center, which is operated by the Smithsonian Astrophysical Observatory for and on behalf of NASA under contract NAS8-03060. 
JLB received funding from the European Union (ERC, MAGNIFY, Project 101126182).
The authors have no relevant financial or non-financial interests to disclose. 

All authors contributed to the study conception and design. Literature review on \gc\ was performed by YN and on CVs by JLB. RGR conducted the simulations and analyzed them, contributed all text on simulation methodology, results, and analysis, and provided detailed edits to the remaining sections. CE contributed text and expertise related to UV and optical spectropolarimetry, and provided detailed comments which shaped the structure of the manuscript. CEJ provided feedback on the simulations and revised the first draft of the manuscript to ensure cohesiveness. All authors commented on previous versions of the manuscript, and have read and approved the final manuscript.

Ethics declaration: not applicable.

\end{document}